\documentclass[11pt]{article}
\usepackage{epsfig}
\usepackage{amssymb}
\newcommand{\sect}[1]{ \section{#1} \setcounter{equation}{0} }

\newcommand{\Dslash}{D \! \! \! \! /} 
 
\newcommand{\pslash}{p \! \! \! /}

\newcommand{\partialslash}{\partial \! \! \! /} 
\newcommand{\half}{\mbox{\small{$\frac{1}{2}$}}} 
\newcommand{\third}{\mbox{\small{$\frac{1}{3}$}}} 
 
\newcommand{\pitwo}{\mbox{\small{$\frac{\pi}{2}$}}} 
\newcommand{\pisix}{\mbox{\small{$\frac{\pi}{6}$}}} 
\newcommand{\MSbar}{\overline{\mbox{MS}}}

\newcommand{\Nc}{N_{\!c}}
\newcommand{\Nf}{N_{\!f}}
\newcommand{\NF}{N_{\!F}}
\newcommand{\NA}{N_{\!A}}

\begin{document}
\title{Six dimensional QCD at two loops}
\author{J.A. Gracey, \\ Theoretical Physics Division, \\ 
Department of Mathematical Sciences, \\ University of Liverpool, \\ P.O. Box 
147, \\ Liverpool, \\ L69 3BX, \\ United Kingdom.} 
\date{} 
\maketitle 

\vspace{5cm} 
\noindent 
{\bf Abstract.} We construct the six dimensional Quantum Chromodynamics (QCD) 
Lagrangian in a linear covariant gauge and subsequently renormalize it at two
loops in the $\MSbar$ scheme. The coupling constant corresponding to the gauge
interaction is asymptotically free for all numbers of quark fields, $\Nf$. 
Analysing the $\beta$-functions yields a rich spectrum of fixed points. For 
instance, the conformal window in the six dimensional theory is at 
$\Nf$~$=$~$16$ for the $SU(3)$ colour group. The critical theory structure is 
similar to that of an $O(N)$ scalar theory in eight dimensions. Using the large
$N$ expansion the latter is shown to be in the same universality class as the 
Heisenberg ferromagnet. Similarly using the large $\Nf$ expansion, six 
dimensional QCD is shown to be in the same class as the two dimensional 
non-abelian Thirring model and four dimensional QCD. Abelian gauge theories are
also renormalized at high loops in six and eight dimensions. It is shown that 
the gauge parameter only appears in the electron anomalous dimension at one 
loop, similar to four dimensions.

\vspace{-18.5cm}
\hspace{13cm}
{\bf LTH 1072}

\newpage

\sect{Introduction.}

In recent years there has been renewed interest in the properties of higher 
dimensional quantum field theories. This has been in part due to the long
established fact that Quantum Chromodynamics (QCD) has a non-trivial fixed
point in strictly four spacetime dimensions for a range of values of the number
of quarks, $\Nf$, \cite{1}. Known as the conformal window the range is
$9$~$\leq$~$\Nf$~$\leq$~$16$ for the $SU(3)$ colour group. It is due to the two
loop $\beta$-function having a non-trivial fixed point when the sign of the one
loop and two loop terms are different, \cite{1}. Termed as the Banks-Zaks fixed
point the lower bound of the window is determined by the two loop 
$\beta$-function coefficient. However, it is not clear whether this is the 
actual range of the window since the value of the coupling constant at the 
lower end is beyond the conventional range for perturbative reliability. Other 
non-abelian gauge theories, such as those with supersymmetry, have a similar 
property. While the conformal window of QCD was the first to be studied, the 
current vision is that similar fixed points in gauge theories with other group 
symmetry could give insight into the theory believed to lie beyond the Standard
Model. For instance, operators which are not relevant at the Gaussian fixed
point could become relevant at a non-trivial fixed point and hence drive the
dynamics. Recent analyses and refinements of fixed point locations for various
colour groups and representations can be found, for example, in 
\cite{2,3,4,5,6,7,8,9,10}. One key to this is the extension of our 
understanding of conformal field theory in two spacetime dimensions to 
dimensions greater than two. This is not a trivial task as the conformal group 
in the former dimension is infinite dimensional but finite beyond two spacetime
dimensions. One notable property of two dimensions was the $c$-theorem, 
\cite{11}, which carries information on the renormalization group flow of a 
theory. There have been attempts to find its $d$-dimensional generalization 
such as the $a$-theorem, \cite{12}. The aim there is to find the function, 
similar to two dimensions, which is positive in the renormalization group flow 
from ultraviolet to infrared. 

Parallel to this analysis, and in parts underlying it, is the need to determine
and study the explicit renormalization group functions of field theories in
dimensions greater than four. There has been work in various directions
recently. In particular six dimensional $O(N)$ scalar $\phi^3$ theory has
received detailed attention, \cite{13,14,15,16,17}. These perturbative studies 
are complementary to the modern application of the conformal bootstrap 
programme, originally developed in \cite{18,19,20,21,22,23,24} and extended in
\cite{25,26,27,28}. Indeed one motivation in \cite{29,30,31} was the connection
of this cubic scalar theory with higher spin theories which naturally emerge 
from AdS/CFT's. In \cite{15,17} the conformal window was established in 
$d$~$=$~$6$~$-$~$2\epsilon$ dimensions, extending the one loop result of
\cite{32}, and the spectrum of fixed points determined to three loops. This was
later extended to four loops in \cite{33}. The reason why the window was 
studied away from the critical dimension of $\phi^3$ theory is that, in 
principle, it ought to be possible to connect the fixed points in the higher 
dimensional theory with conformal field theories in lower dimensions including 
two. The latter is important as conformal field theories have been classified 
there. An example of this ambition was given in \cite{33}. There using 
summation approaches, which are standard in condensed matter theory, various 
critical exponents derived in the $\epsilon$ expansion of the six dimensional 
theory were summed to access the discrete dimensions lower than six. Central to
this was the knowledge of the value of the corresponding critical exponent in 
the underlying two dimensional conformal field theory. Using this as a boundary
condition for the four loop Pad\'{e} approximant, estimates for the exponents 
were found to be competitive with strong coupling methods for models of 
percolation, for instance. 
 
This connection across the dimensions is not a novel observation as it dates
from the work of Wilson and Fisher, \cite{34}. They observed that in 
$d$-dimensions, where one can regard $d$~$=$~$D$~$-$~$2\epsilon$, with $D$ 
integer and the critical dimension of a theory, different quantum field 
theories can have the same critical exponents. This equivalence occurs at the 
non-trivial $d$-dimensional fixed point of the respective $\beta$-functions 
which is now termed the Wilson-Fisher fixed point. This property, known as
universality, is a powerful computational tool for analysing quantum field 
theories. The most common example is the relation between the two dimensional 
$O(N)$ nonlinear $\sigma$ model and four dimensional $O(N)$ $\phi^4$ theory, 
\cite{34}. Each is perturbatively renormalizable in their critical dimensions 
but at the $d$-dimensional Wilson-Fisher fixed point they are in the same 
universality class. That they can be seen to be connected across the dimensions
is possible through the large $N$ expansion where $1/N$ plays the role of a 
dimensionless coupling constant in $d$-dimensions. Thus the apparently 
perturbatively nonrenormalizable nonlinear $\sigma$ model is non-perturbatively
renormalizable in $d$-dimensions in the large $N$ expansion. To see this 
equivalence in depth is possible through the work of Vasiliev's group, 
\cite{35,36,37}. In \cite{35,36,37} the critical exponents of the basic fields 
and operators were determined to the third term as functions of $d$. This is 
$O(1/N^3)$ for the matter field anomalous dimension and $O(1/N^2)$ for what 
would be the force or bound state field. The exponents for the $\beta$-function
slopes of the respective models are known at $O(1/N^2)$ in \cite{36,38}. When
these critical exponents are expanded using $d$~$=$~$D$~$-$~$2\epsilon$ 
relative to the respective underlying theories, can one then appreciate the 
exact agreement with the explicit perturbative renormalization group functions.
This includes, for instance, recent six loop $\MSbar$ computations of the field
wave function anomalous dimension in four dimensional $O(N)$ $\phi^4$ theory, 
\cite{39}.

What has been established more recently is the extension of this Wilson-Fisher
fixed point universality chain to six dimensions in \cite{17,33}. Thus one
natural question, which has been posed in several articles \cite{15,17,40},
concerns whether there is a tower of such theories and if so what is the
algorithm to construct each in a specific spacetime dimension. Part of this 
article addresses this since we construct an eight dimensional $O(N)$ scalar 
field theory which we will show is in the same universality class as that of 
$O(N)$ scalar theories. It transpires that the process to build the tower is 
straightforward. In essence it is in keeping with the vision of Wilson that the
universal theory is an infinite set of (local) operators, obeying a symmetry 
such as $O(N)$, which become relevant in the renormalization group sense in the
critical dimension. Otherwise such operators are irrelevant in other critical 
dimensions. These remarks have to be qualified by noting that they correspond 
to massless theories. If mass parameters are permitted then relevant operators 
of theories with lower critical dimensions will be part of the universal 
Lagrangian. We will briefly study the massive extension of our eight 
dimensional $O(N)$ theory too as it will transpire that this together with the 
massless version have structural similarities with the second and main thread 
of this article. This is the application of the above ideas to non-abelian 
gauge theories with the intention of determining connections of Lagrangians of 
spin-$1$ fields across $d$-dimensions. 

In principle the construction of a similar tower of gauge theories should be
feasible based on what has been found in the scalar theory case. Moreover, it
should be relevant to possible directions beyond the Standard Model. For 
instance, for certain gauge groups, such as
$SU(3)$~$\times$~$SU(2)$~$\times$~$U(1)$, there may be a flow to a non-trivial 
fixed point which connects with a unified theory. Also understanding the low 
energy dynamics of Yang-Mills theories is currently a major goal. While the 
canonical QCD Lagrangian more than adequately describes high energy quark and 
gluon dynamics, it lacks many features in the infrared. One notable problem is 
that quarks and gluons have fundamental massless propagators, which derive from
the Lagrangian, but these contradict the fact that these quanta are confined
and not observed in nature. In other words operators which are ultraviolet
irrelevant may become infrared relevant and dominate the infrared dynamics to 
the extent that the quark and gluon propagators cease having their fundamental 
form. One such operator which has received attention at various times is the 
purely gluonic dimension six operator  
$f^{abc} G_{\mu\nu}^a \, G^{b \, \mu\sigma} \, G^{c \,\nu}_{~~\,\sigma}$ where
$G_{\mu\nu}^a$ is the gluon field strength and $f^{abc}$ are the colour group
structure constants. Clearly this operator is perturbatively nonrenormalizable 
in four dimensions. However, based on the scalar theory picture it is possible 
to consider it in a renormalizable six dimensional Lagrangian. If the fixed 
point structure of the higher theory is such that the operator's coupling 
becomes relevant through the $\epsilon$-expansion in four dimensions at a 
non-trivial point then it could be part of the structure governing the infrared
dynamics of gluons. While we have highlighted this specific operator, we 
acknowledge that there are likely to be many other operators with higher 
dimensions. However, it is worth considering the simplest extension of the 
Wilson vision for a universal gauge theory. As an aside six dimensional gauge 
theories have received attention at various times, \cite{41,42,43,44,45,46,47}.
For instance, in \cite{41} a version of six dimensional QCD, similar to what we
will consider here, was studied at one loop but in the background field gauge. 
The motivation was in part to give insight into supersymmetric extensions and 
to provide a framework to connect with string dynamics. An approach along 
similar grounds but motivated by a model building framework can be found in
\cite{42}. Partly related to these is a second area of attention which is the 
explicit examination of six dimensional supersymmetric theories. While we do 
not consider supersymmetry explicitly here the Lagrangians of the 
supersymmetric gauge theories, \cite{43,47}, have similarities to our 
non-supersymmetric one. 

Therefore, our goal will be to construct the perturbatively renormalizable six 
dimensional non-abelian gauge theory and compute its renormalization group 
functions to two loops in the $\MSbar$ scheme. We have to proceed to this order
as it will be apparent that the one loop or leading order is effectively 
trivial from the fixed point structure point of view. From the computational 
side a two loop renormalization provides a highly non-trivial check on the 
explicit construction such as the issue of the gauge fixing in six dimensions. 
Related to this is the check that the $\MSbar$ $\beta$-functions have to be 
independent of the linear covariant gauge fixing parameter. We will show this 
separately for each of the three $3$-point vertices. Concerning the aim of
connecting with gluon infrared dynamics in four dimensions, it will turn out 
that like the eight dimensional $O(N)$ scalar theory the six dimensional gluon 
propagator will have a double pole propagator. In four dimensions such a 
propagator was believed for a while, \cite{48}, to be the form in the infrared 
which ensured a linearly rising interquark potential and hence the confining 
force. However, current Landau gauge lattice measurements and Schwinger-Dyson 
studies of the gluon propagator in four dimensions suggest otherwise in that 
the propagator freezes to a finite non-zero value at zero momentum. See, for 
instance,  \cite{49,50,51,52,53,54,55,56,57,58,59}. However, we will also 
provide modified gluon and Faddeev-Popov ghost propagators which closely 
resemble those developed and used in four dimensional models of the infrared. 
Our approach is via corrections to scaling and is offered as a novel but 
alternative insight into such models rather than a justification. One further 
remark needs to be made in relation to the earlier scalar theory discussion. It
concerns the problem of which theories lie in the tower of gauge theories. It 
transpires that a similar chain has been known for some time in the dimension 
range $2$~$<$~$d$~$<$~$4$. In \cite{60} it was shown that the two dimensional 
non-abelian Thirring model and four dimensional QCD were connected in 
$d$-dimensions at their Wilson-Fisher fixed points. This was accessed via the 
large $\Nf$ expansion where it is the number of quark flavours, $\Nf$, which is
the dimensionless parameter and not the number of colours. Indeed this 
$d$-dimensional equivalence was exploited in, for example, \cite{61,62,63} in 
order to determine various large $\Nf$ critical exponents as functions of $d$. 
While not computed to as high an order in powers of $1/\Nf$ as the scalar field
theories, these $d$-dimensional exponents will play a very useful role in 
connecting to and establishing the six dimensional gauge theory as being part 
of the tower with the non-abelian Thirring model as its foundation stone. Once
this connection has been achieved we will be able to study the questions of the 
existence of a conformal window and the two loop fixed point structure. As a 
corollary and as a stepping stone beyond six dimensions we will specialize to 
six and eight dimensional abelian gauge theories due to recent interest in
these theories, \cite{64,65}. This will be at three and two loops respectively.
Again we will establish the connectivity in the tower of $d$-dimensional 
abelian gauge theories living at the Wilson-Fisher fixed point. The main 
motivation for this is as a prelude to analysing an eight dimensional 
non-abelian extension. That is a more involved exercise for a later article,
since one has to determine the set of independent non-abelian operators which 
build the perturbatively renormalizable Lagrangian. Some work on eight 
dimensional operators has been provided in \cite{66} but this analysis was not 
motivated for constructing a Lagrangian. Rather it was for ascertaining the 
basis for the operator product expansion and QCD sum rules.

The article is organized as follows. The algorithm to construct the candidate
higher dimensional scalar theories as well as the notation we will use is
discussed in section $2$. The eight dimensional $O(N)$ scalar theory is 
renormalized in the subsequent section and the fixed point structure analysed 
after showing that it is in the same universality class as the Heisenberg 
ferromagnet. A corollary of that computation is to consider the $Sp(N)$ version 
in section $4$. The focus then changes to gauge theories and the construction
of the higher dimensional gauge theories is discussed in section $5$. In order
to make the Wilson-Fisher fixed point connection relevant results from the 
large $\Nf$ expansion are provided in section $6$. The analysis of the two loop
renormalization group functions of six dimensional QCD is given in the 
following section. Subsequently, we specialize to abelian gauge theories in
section $8$ before providing a concluding viewpoint in section $9$. An appendix
records values for various eight dimensional one and two loop integrals. 

\sect{Background.}

We begin by discussing the construction of the higher dimensional $O(N)$ scalar 
quantum field theories which lie in the same 
universality class as the two dimensional nonlinear $\sigma$ model and $\phi^4$
theory in four dimensions at the Wilson-Fisher fixed point. These theories are 
equivalent in $2$~$<$~$d$~$<$~$4$ dimensions which can be seen within the large
$N$ expansion. In \cite{17} the extension of the chain to six dimensions was
analysed in full and moreover gives a clue as to how to extend the sequence to
eight and higher dimensions. To appreciate this it is instructive to consider
the two lowest dimensional Lagrangians which are 
\begin{equation}
L_\phi^{(2)} ~=~ \frac{1}{2} \partial_\mu \phi^i \partial^\mu \phi^i ~+~ 
\frac{1}{2} g_1 \sigma \phi^i \phi^i ~-~ \frac{1}{2} \sigma 
\label{lag2}
\end{equation}
for the nonlinear $\sigma$ model and
\begin{equation}
L_\phi^{(4)} ~=~ \frac{1}{2} \partial_\mu \phi^i \partial^\mu \phi^i ~+~ 
\frac{1}{2} g_1 \sigma \phi^i \phi^i ~+~ \frac{1}{2} \sigma^2 
\label{lag4}
\end{equation}
for the four dimensional quartic theory. In (\ref{lag2}) the field $\sigma$
plays the role of a Lagrange multiplier field as ordinarily one would not have
a linear term in a Lagrangian. The multiplier is necessary in order to restrict
the $O(N)$ scalar fields to lie on the $N$-sphere. Choosing a coordinate 
system for the constraint that the length of $\phi^i$ is fixed to be the
coupling constant would produce the nonlinear version of (\ref{lag2}) which is 
\begin{equation}
L_\phi^{(2)} ~=~ \frac{1}{2} g_{ab}(\phi) \partial_\mu \phi^a \partial^\mu 
\phi^b ~.
\label{lagnlsm2}
\end{equation}
Here $1$~$\leq$~$a$~$\leq$~$(N-1)$ and $g_{ab}(\phi)$ is the metric of the 
sphere in the chosen coordinate system. Equally we have not expressed 
(\ref{lag4}) in its canonical form as there the $\sigma$ field is regarded as
an auxiliary field. Eliminating it produces 
\begin{equation}
L_\phi^{(4)} ~=~ \frac{1}{2} \partial_\mu \phi^i \partial^\mu \phi^i ~-~ 
\frac{g_1^2}{8} \left( \phi^i \phi^i \right)^2 
\label{lagphi4}
\end{equation}
where the quartic interaction is apparent. While both (\ref{lagnlsm2}) and
(\ref{lagphi4}) are the usual formulations it is (\ref{lag2}) and (\ref{lag4})
which best indicate that they both lie in the same universality class. This is
because both have a common interaction. The only differences are in the terms 
involving $\sigma$. The key point is that the coupling constant $g_1$ has 
different canonical dimensions in each Lagrangian and this is as a result of 
these $\sigma$ dependent terms. They define the dimension of each coupling 
which can be seen if one rescales $\sigma$~$\to$~$\sigma/g_1$. Indeed that is 
the version used in the critical point large $N$ method of \cite{35,36,37}. In 
other words the commonality of the $\sigma \phi^i \phi^i$ interaction is what 
determines the universality. 

This is evident in the extension of \cite{15,17} to six dimensions as that 
Lagrangian is  
\begin{equation}
L_\phi^{(6)} ~=~ \frac{1}{2} \partial_\mu \phi^i \partial^\mu \phi^i ~+~ 
\frac{1}{2} \partial_\mu \sigma \partial^\mu \sigma ~+~
\frac{1}{2} g_1 \sigma \phi^i \phi^i ~+~
\frac{1}{6} g_2 \sigma^3 ~.
\label{lag6}
\end{equation}
The relation to (\ref{lag2}) and (\ref{lag4}) is that it has the same common
interaction as before but the $\sigma$ dependent term no longer contributes to
the free Lagrangian. The reason for this new interaction is primarily to ensure
the six dimensional Lagrangian is perturbatively renormalizable. One 
consequence is that there is a vector of $\beta$-functions and hence a rich 
fixed point structure emerges, \cite{17}. However, it has been checked that the
renormalization group functions at three and four loops, \cite{17,33}, can be 
converted into critical exponents in the large $N$ expansion which agree 
precisely with the exponents computed directly in $1/N$, \cite{35,36,37,38}. 
This is a non-trivial observation since the cubic $\sigma$ interaction with the
additional coupling constant plays a key role in ensuring consistency. For this
article we will term this and similar additional interactions in other theories
as the spectator interactions. This is because the connecting interaction, 
$\sigma \phi^i \phi^i$, is central to the universality and the spectators are 
dimension dependent. Moreover, when we extend the picture to gauge theories 
this connecting interaction actually connects the quantum of the underlying 
force with matter. One difference (\ref{lag6}) has with the other lower 
dimensional theories is that $\sigma$ cannot be eliminated either as a Lagrange 
multiplier or an auxiliary field. Once the connectivity of (\ref{lag2}), 
(\ref{lag4}) and (\ref{lag6}) has been established it will be apparent how one 
extends the tower of Lagrangians to higher dimensions. There are several key 
ingredients. One is the connecting $\sigma \phi^i \phi^i$ interaction and the
second is that the theory has to be perturbatively renormalizable in the higher
dimension. In addition the field $\phi^i$ has, of course, to lie in the same 
symmetry group which is a minor observation. Given this it is straightforward 
to write down a candidate eight dimensional Lagrangian for the equivalence at 
the Wilson-Fisher fixed point which is 
\begin{equation}
L_\phi^{(8)} ~=~ \frac{1}{2} \partial_\mu \phi^i \partial^\mu \phi^i ~+~ 
\frac{1}{2} \left( \Box \sigma \right)^2 ~+~ 
\frac{1}{2} g_1 \sigma \phi^i \phi^i ~+~
\frac{1}{6} g_2 \sigma^2 \Box \sigma ~+~ \frac{1}{24} g_3^2 \sigma^4 ~.
\label{lag8}
\end{equation}
We have normalized the coupling constants in each interaction so that the
Feynman rule for each is effectively unity. A similar pattern is present with 
the other three theories in that the $\sigma$ dependent term is extended to a 
quartic one as might be expected. However, contained within the perturbative 
renormalizability criterion is the understanding that one has a set of 
independent operators with which to formulate the Lagrangian. This is the 
reason for an interaction with a derivative coupling. On dimensional grounds 
there are more possible interactions with derivatives but only one is 
independent. They are all related by integration by parts where total 
derivative operators can be dropped from the Lagrangian as they can be 
integrated out of the action. The other major difference which first appears 
here is the presence of a double pole $\sigma$ propagator. This is due to the 
fact that the canonical dimension of $\sigma$ at the Wilson-Fisher fixed point 
is always $2$ which is why $\sigma$ has a momentum dependent propagator in 
(\ref{lag6}) but not in lower dimensions. It will turn out that in the gauge 
theory context a similar higher order pole propagator will emerge but in a 
lower dimension. So (\ref{lag8}) could be regarded as a simple 
laboratory for testing ideas in higher dimensional QCD in much the same way 
that six dimensional $\phi^3$ theory was once regarded as a test bed for four 
dimensional QCD, \cite{67,68}, partly due to both being asympotically free.  

\sect{Eight dimensional $O(N)$ scalar theory.}

While we have formulated a candidate eight dimensional scalar theory according
to certain criteria we still have to test out whether the renormalization group
functions of (\ref{lag8}) are consistent with the large $N$ critical exponents.
To do this at a credible and non-trivial level requires a two loop analysis. 
Therefore, we have constucted the anomalous dimensions of the two fields and 
the $\beta$-functions to the requisite orders. Specifically we have determined 
the anomalous dimensions of $\phi^i$ and $\sigma$ at three loops and the 
$\beta$-functions of $g_1$ and $g_2$ at two loops. For the coupling of the 
quartic term we have computed $\beta_3(g_1,g_2,g_3)$ at one loop. The reason 
for the different loop orders stems partly from computational constraints. For 
instance, the field anomalous dimensions do not have any $g_3$ dependence at 
one loop. So that the two loop term of $\beta_3(g_1,g_2,g_3)$ will not have any
effect on the checks with the large $N$ exponents. However, we have evaluated 
the wave function renormalization at three loops so that we can in fact check 
that the two loop renormalization is consistent. This is because the triple and
double poles in $\epsilon$ in the three loop renormalization constants are 
determined by lower loop information. Here $\epsilon$ is the regularizing 
parameter in dimensional regularization which we use throughout. Hence it is 
possible to check the result is consistent with this property of the 
renormalization group equation. One technical limitation arises in the 
renormalization of the vertices. In \cite{33} we were able to exploit a 
property of six dimensions which was that a propagator of the form $1/(k^2)^2$,
where $k$ is the momentum, does not introduce spurious infrared infinities. 
Therefore, one could renormalize the $3$-point vertices by nullifying one 
external momentum. This simple infrared rearrangement meant that the vertex 
renormalization devolved to a problem of evaluating $2$-point Feynman graphs 
which is computationally much simpler than a full $3$-point function, 
\cite{33}. For (\ref{lag8}) this nullification of the momentum on an external 
leg of a $3$-point vertex cannot be used. The main reason for this is that the 
$\sigma$ propagator is itself now $1/(k^2)^2$. In eight dimensions this on its 
own does not introduce any infrared problems. However, if an external momentum 
is nullified then Feynman integrals will have factors such as $1/(k^2)^4$ which
will produce unwanted infrared infinities which cannot be disentangled from the
desired ultraviolet one. Therefore, for the $3$-point vertex renormalization we
have chosen to evaluate the Feynman integrals for the case when none of the 
external momenta are nullified. Moreover, we will carry out the subtraction of 
infinities in the $\MSbar$ scheme at the fully symmetric point where the 
squared external momenta are all equal to $(-\tilde{\mu}^2)$ where 
$\tilde{\mu}$ is the mass scale introduced to ensure the coupling constants are
dimensionless in $d$-dimensions. One benefit of considering the symmetric point
is that it corresponds to a non-exceptional momentum configuration. So there 
are no infrared issues and the poles in $\epsilon$ which emerge are purely 
ultraviolet. For the $4$-point vertex the same issues arise. One cannot nullify
an external momentum to reduce the computation to a $3$-point one as then the 
momentum configuration is exceptional. Therefore, we have chosen to compute the
one loop $4$-point function at its fully symmetric point to ensure the result 
is infrared safe. 

Having outlined the general method of computing the renormalization group
functions we now discuss the more practical technical aspects of the process.
This approach described here was also applied to the gauge theory computations 
presented later. Our calculations were carried out automatically using
symbolic manipulation programmes written in the language {\sc Form}, 
\cite{69,70}. The initial part of this is to generate all the Feynman diagrams
electronically with the {\sc Qgraf} package, \cite{71}. Once this is achieved 
the graphs are individually passed to an integration routine. The final stage 
of the process is to sum all the graphs and extract the renormalization 
constants. This latter part is achieved automatically by using the algorithm of 
\cite{72}. In essence one computes each graph as a function of the bare
parameters. These are the three coupling constants in (\ref{lag8}) and in the
case of the gauge theories the gauge parameter. The renormalized variables are
introduced by rescaling with the respective renormalization constants 
corresponding to the constant of proportionality. This in effect introduces the
counterterms automatically and bypasses the need to carry out subtractions on
each individual graph which would be tedious for a high loop analysis. The bulk
of the work is in the integration routine and for each of the three types of
Green's functions, $2$-, $3$- and $4$-point, we have used the Laporta 
algorithm, \cite{73}. This is an elegant technique which systematically creates
all the relations between scalar Feynman integrals using integration by parts 
and then algebraically solves them in terms of a base set of integrals. This 
set is known as the master integrals and is ordinarily a relatively small set. 
They are evaluated directly if, for example, they are nested bubble graphs, or 
by non-integration by parts methods. The version of the Laporta algorithm we 
used was {\sc Reduze}, \cite{74,75}. It creates a database of relations from 
which we extract the required integrals for each Green's function in {\sc Form}
notation and then include the relations as a {\sc Form} module in the automatic
computation. For (\ref{lag8}) {\sc Reduze} is particularly appropriate for the 
$2$- and $3$-point functions since the higher pole $\sigma$ propagator requires
a larger order of integration than is ordinarily the case. The final stage is 
the substitution of the expressions for the master integrals. As we are 
interested in the structure of (\ref{lag8}) we have to determine master 
integrals in eight dimensions. This is more straightforward than may initially 
seem which is due to the fact that the relevant masters are already known in 
lower dimensions. One can connect with these results via the Tarasov method, 
\cite{76,77}, which allows one to relate $d$-dimensional Feynman integrals with
integrals in $(d+2)$-dimensions. The latter have the same topology as the lower
dimensional one but with powers of propagator which are larger than those of 
the original. However, such integrals can be reduced to the corresponding 
master in the higher dimension by application of the Laporta algorithm. 
Therefore, one can readily construct relations between masters in different 
dimensions plus lower level masters which are already available. Therefore, if 
a lower dimensional master is available it can be used to immediately determine
the value of the corresponding master in two dimensions higher. This process 
was used to deduce the four loop masters for the $2$-point functions in six 
dimensions in \cite{33}. To aid an interested reader we have presented all the 
relevant two loop eight dimensional masters for the $3$-point function at the 
fully symmetric point to various orders in $\epsilon$, where 
$d$~$=$~$8$~$-$~$2\epsilon$, in Appendix A as well as the one loop 
$4$-point box at its symmetric point. The former complement the same values for
the six dimensional case which were presented in \cite{78}. 

Applying this procedure we have found the various renormalization group 
functions for (\ref{lag8}) are 
\begin{eqnarray}
\gamma_\phi(g_1,g_2,g_3) &=& -~ \frac{g_1^2}{12} 
+ \left[ 801 N g_1^2 + 2250 g_1^2 - 10800 g_1 g_2 - 1072 g_2^2 \right]
\frac{g_1^2}{777600} \nonumber \\
&& +~ \left[ 179415 N^2 g_1^4 + 89688870 N g_1^4 + 419904000 \zeta_3 g_1^4
- 520870500 g_1^4 \right. \nonumber \\
&& \left. ~~~~+ 56945160 N g_1^3 g_2 + 87750000 g_1^3 g_2 
+ 3280536 N g_1^2 g_2^2 + 186624000 \zeta_3 g_1^2 g_2^2 \right. \nonumber \\
&& \left. ~~~~- 491324400 g_1^2 g_2^2 - 116640000 g_1^2 g_3^2 
- 65550960 g_1 g_2^3 - 90720000 g_1 g_2 g_3^2 \right. \nonumber \\
&& \left. ~~~~- 437392 g_2^4 - 11275200 g_2^2 g_3^2 - 5370300 g_3^4 \right] 
\frac{g_1^2}{50388480000} ~+~ O(g_i^8) \nonumber \\
\gamma_\sigma(g_1,g_2,g_3) &=& \left[ 9 N g_1^2 - 8 g_2^2 \right] 
\frac{1}{1080} \nonumber \\
&& +~ \left[ -~ 5265 N g_1^4 + 51840 N g_1^3 g_2 + 4392 N g_1^2 g_2^2 
- 2344 g_2^4 - 21600 g_2^2 g_3^2 \right. \nonumber \\
&& \left. ~~~~- 12150 g_3^4 \right] \frac{1}{17496000} \nonumber \\
&& +~ \left[ - 255817035 N^2 g_1^6 - 944784000 \zeta_3 N g_1^6 
+ 1761646725 N g_1^6 \right. \nonumber \\
&& \left. ~~~~- 47764080 N^2 g_1^5 g_2 - 66703500 N g_1^5 g_2 
- 85536 N^2 g_1^4 g_2^2 \right. \nonumber \\
&& \left. ~~~~- 1049760000 \zeta_3 N g_1^4 g_2^2 + 3348988740 N g_1^4 g_2^2 
+ 1043199000 N g_1^4 g_3^2 \right. \nonumber \\
&& \left. ~~~~+ 370958940 N g_1^3 g_2^3 + 660474000 N g_1^3 g_2 g_3^2 
+ 19729152 N g_1^2 g_2^4 \right. \nonumber \\
&& \left. ~~~~+ 80578800 N g_1^2 g_2^2 g_3^2 + 28048275 N g_1^2 g_3^4 
+ 36288000 \zeta_3 g_2^6 - 376270760 g_2^6 \right. \nonumber \\
&& \left. ~~~~+ 445780800 g_2^4 g_3^2 + 389431800 g_2^2 g_3^4 
+ 75451500 g_3^6 \right] \frac{1}{1133740800000} \nonumber \\
&& +~ O(g_i^8) \nonumber \\
\beta_1(g_1,g_2,g_3) &=& \left[ 9 N g_1^2 + 180 g_1^2 - 240 g_1 g_2 - 8 g_2^2 
\right] \frac{g_1}{2160} \nonumber \\
&& +~ \left[ 187920 N g_1^4 - 516375 g_1^4 + 65880 N g_1^3 g_2 
+ 486000 g_1^3 g_2 + 2196 N g_1^2 g_2^2 \right. \nonumber \\
&& \left. ~~~~- 827280 g_1^2 g_2^2 - 729000 g_1^2 g_3^2 - 36120 g_1 g_2^3 
- 162000 g_1 g_2 g_3^2 - 1172 g_2^4 \right. \nonumber \\
&& \left. ~~~~- 10800 g_2^2 g_3^2 - 6075 g_3^4 \right] 
\frac{g_1}{17496000} ~+~ O(g_i^7) \nonumber \\
\beta_2(g_1,g_2,g_3) &=&
\left[ 270 N g_1^3 + 27 N g_1^2 g_2 + 76 g_2^3 \right] \frac{1}{2160} 
\nonumber \\
&& +~ \left[ - 91125 N g_1^5 + 662985 N g_1^4 g_2 + 14715 N g_1^3 g_2^2 
+ 121500 N g_1^3 g_3^2 \right. \nonumber \\
&& \left. ~~~~- 7083 N g_1^2 g_2^3 + 8100 N g_1^2 g_2 g_3^2 - 43394 g_2^5 
+ 43200 g_2^3 g_3^2 \right. \nonumber \\
&& \left. ~~~~+ 109350 g_2 g_3^4 \right] \frac{1}{11664000} ~+~ O(g_i^7) 
\nonumber \\
\beta_3(g_1,g_2,g_3) &=& \left[ 810 N g_1^4 + 27 N g_1^2 g_3^2 + 160 g_2^4 
+ 696 g_2^2 g_3^2 + 405 g_3^4 \right] \frac{1}{1620} ~+~ O(g_i^6) 
\label{rge8}
\end{eqnarray}
where $\zeta_z$ is the Riemann zeta function and the order symbol in
perturbative expressions throughout indicates any combination of couplings 
whose powers sum to that indicated. While the three loop field anomalous 
dimensions satisfy internal consistency checks the main motivation is to 
ascertain whether (\ref{lag8}) is in the same Wilson-Fisher universality class 
as (\ref{lag2}), (\ref{lag4}) and (\ref{lag6}) which requires computing the 
critical exponents in the large $N$ expansion. To do this we follow the same 
prescription and method introduced in \cite{17}. First, we introduce rescaled 
coupling constants 
\begin{equation}
g_1 ~=~ \sqrt{\frac{120\epsilon}{N}} x ~~~,~~~
g_2 ~=~ \sqrt{\frac{120\epsilon}{N}} y ~~~,~~~
g_3^2 ~=~ \frac{120\epsilon}{N} z ~.
\end{equation}
Then solving $\beta_i(g_1,g_2,g_3)$~$=$~$0$ we find 
\begin{eqnarray}
x &=& 1 ~+~ \left[ - 110 + 252 \epsilon \right] \frac{1}{N} ~+~ 
\left[ 18150 - 48475 \epsilon \right] \frac{1}{N^2} ~+~ 
\left[ 67232500 - 551223750 \epsilon \right] \frac{1}{N^3} \nonumber \\
&& +~ O \left( \epsilon^2 ; \frac{1}{N^4} \right) \nonumber \\
y &=& -~ 15 ~+~ \left[ 14250 - 21210 \epsilon \right] \frac{1}{N} ~+~ 
\left[ - 36182250 + 89882625 \epsilon \right] \frac{1}{N^2} \nonumber \\
&& +~ \left[ 128836402500 - 416828165250 \epsilon \right] \frac{1}{N^3} ~+~ 
O \left( \epsilon^2 ; \frac{1}{N^4} \right) \nonumber \\
z &=& -~ 60 ~-~ \frac{12000}{N} ~+~ \frac{1045416000}{N^2} ~-~
\frac{13222012800000}{N^3} ~+~
O \left( 1 ; \frac{1}{N^4} \right) 
\end{eqnarray}
where the order symbol for $1/N$ expansions indicates the truncation powers of
both expansions. Using these to evaluate $\gamma_\phi(g_1,g_2,g_3)$ and
$\gamma_\sigma(g_1,g_2,g_3)$ at this large $N$ fixed point we find exponents 
are
\begin{eqnarray}
\eta &=& \left[ -~ 20 \epsilon + \frac{89}{3} \epsilon^2 \right] 
\frac{1}{N} ~+~ 
\left[ 4400 \epsilon - \frac{77950}{3} \epsilon^2 \right] \frac{1}{N^2} ~+~ 
\left[ -~ 968000 \epsilon + 18577400 \epsilon^2 \right] \frac{1}{N^3}
\nonumber \\
&& +~ O \left( \epsilon^3 ; \frac{1}{N^4} \right) \nonumber \\
\eta ~+~ \chi &=& \epsilon ~+~ 
\left[ -~ 420 \epsilon + 673 \epsilon^2 \right] \frac{1}{N} ~+~ 
\left[ 428400 \epsilon - \frac{4544450}{3} \epsilon^2 \right] \frac{1}{N^2} ~+~
O \left( \epsilon^3 ; \frac{1}{N^3} \right)
\label{largen8}
\end{eqnarray}
for comparison with the exponents given in \cite{35,36,37} when expanded in 
powers of $\epsilon$ where $d$~$=$~$8$~$-$~$2\epsilon$. Here $\eta$ relates to 
the renormalization group function $\gamma_\phi(g_1,g_2,g_3)$ and 
$\eta$~$+$~$\chi$ is the exponent underlying $\gamma_\sigma(g_1,g_2,g_3)$ in 
the exponent notation of \cite{35,36}. Comparing the explicit perturbative 
results with the known large $N$ exponents there is precise agreement to 
$O(\epsilon^2)$. While this is not a full proof of the equivalence of 
(\ref{lag8}) with the lower dimensional scalar theories, it has been 
established in a similar way. More importantly it strongly suggests that the 
procedure for constructing a Lagrangian which is a partner in the 
$d$-dimensional tower is well defined. Crucial in this establishment is the 
spectator interactions whose effects first appear at two loops which is a
reason why we constructed the wave function renormalization group functions to 
this order. Given this agreement it is not difficult to write down a candidate 
for the next Lagrangian in the sequence. In ten dimensions following our 
prescription we would have  
\begin{eqnarray}
L_\phi^{(10)} &=& \frac{1}{2} \partial_\mu \phi^i \partial^\mu \phi^i ~+~ 
\frac{1}{2} \left( \Box \partial^\mu \sigma \right) 
\left( \Box \partial_\mu \sigma \right) ~+~ 
\frac{1}{2} g_1 \sigma \phi^i \phi^i ~+~
\frac{1}{6} g_2 \sigma^2 \Box^2 \sigma \nonumber \\
&& +~ \frac{1}{2} g_3 \sigma \left( \Box \sigma \right)^2 ~+~ 
\frac{1}{24} g_4^2 \sigma^3 \Box \sigma ~+~ \frac{1}{120} g_5^3 \sigma^5  
\label{lag10}
\end{eqnarray}
which is renormalizable by power counting. In constructing (\ref{lag10}) we
have ensured that the spectator interactions are independent. Also it shares 
structural similarities to (\ref{lag8}) in that the $\sigma$ propagator has an 
increased pole structure and there are more derivative couplings in addition to
a pure quintic $\sigma$ self-interaction.

We close this section by considering extensions to each of our scalar theories
where (local) operators of dimension lower than the critical dimension for 
renormalizability are included. There are various reasons for this. One is that
the structure of these massive Lagrangians is not unrelated to the Lagrangians
with lower critical dimensions. Indeed this is an indication of the larger 
vision of the operators varying between being relevant and irrelevant in 
different dimensions. A second reason is that one can access an additional 
check on the equivalence of (\ref{lag8}). First, the massive extension of the 
respective four, six and eight dimensional Lagrangians are  
\begin{eqnarray}
L_{\phi\, m}^{(4)} &=& L_\phi^{(4)} ~+~ \frac{1}{2} m_1^2 \phi^i \phi^i 
\nonumber \\
L_{\phi\, m}^{(6)} &=& L_\phi^{(6)} ~+~ \frac{1}{2} m_1^2 \phi^i \phi^i ~+~
\frac{1}{2} m_2^2 \sigma^2 \nonumber \\
L_{\phi\, m}^{(8)} &=& L_\phi^{(8)} ~+~ \frac{1}{2} m_1^2 \phi^i \phi^i ~-~
\frac{1}{2} m_2^2 \sigma \Box \sigma ~+~ \frac{1}{2} m_3^4 \sigma^2 ~+~
\frac{1}{6} m_4^2 \sigma^3 ~. 
\end{eqnarray}
Common to each is a mass term for what one can regard as the matter field 
$\phi^i$. In the context of the large $N$ critical point equivalence, as is 
well known the critical exponent of the $\phi^i$ mass operator is the same as 
$\sigma$ field critical exponent, \cite{35,36}. For the six and eight 
dimensional cases there are additional lower dimensional operators depending 
purely on $\sigma$ including derivative couplings in the latter case. These 
operators are effectively the same as the interactions in theories at lower 
dimensions. When one determines the dimensionality of the associated coupling 
constant in $d$-dimensions then it is clear why these operators are present in 
the massive extensions. Although there is one minor caveat with this in that 
this also includes $2$-point operators which are to be regarded as part of the 
free Lagrangian. So, for instance, the propagators of $L_{\phi\, m}^{(8)}$ are
\begin{equation}
\langle \phi^i(p) \phi^j(-p) \rangle ~=~ 
\frac{\delta^{ij}}{[ p^2 + m_1^2 ]} ~~,~~ 
\langle \sigma(p) \sigma(-p) \rangle ~=~ 
\frac{1}{[(p^2)^2 + m_2^2 p^2 + m_3^4 ]} ~.
\end{equation}
A propagator similar to the denominator of the Stingl propagator, \cite{79},
emerges for the $\sigma$ field. While this may appear to be a non-standard
propagator, it transpires that this form of a propagator can arise in models of
the infrared behaviour of the gluon in QCD. As a further check on our
equivalence we have computed the anomalous dimension of $m_3$ or equivalently
the operator $\sigma^2$ in $L_{\phi\, m}^{(8)}$ at two loops. This is achieved
by inserting the operator in a $\sigma$ $2$-point function but in such a way
that a momentum flows through the operator itself. The reason for this specific
momentum configuration is to ensure that there are no infrared problems in 
extracting the associated operator renormalization constant. Therefore, at two
loops we found  
\begin{eqnarray}
\gamma_{m_3}(g_1,g_2,g_3) &=& 
\left[ - N g_1^2 - 8 g_2^2 - 10 g_3^2 \right] \frac{1}{120} \nonumber \\ 
&& +~ \left[ 14085 N + 24240 N g_1^3 g_2 + 3952 N g_1^2 g_2^2
+ 3330 N g_1^2 g_3^2 \right. \nonumber \\
&& \left. ~~~~ + 18136 g_2^4 + 25640 g_2^2 g_3^2 - 3150 g_3^4 \right] 
\frac{1}{1944000} ~+~ O(g_i^6) ~.
\end{eqnarray}
From this if we evaluate the corresponding critical exponent in the large $N$
expansion using (\ref{largen8}) the exponent is in precise agreement with the
critical exponent $\omega$ computed in \cite{38} at $O(1/N^2)$. In \cite{38}
$\omega$ was determined as it corresponded to the critical slope of the 
$\beta$-function of (\ref{lag4}) and was therefore of interest in accessing the
higher order perturbative structure of the $O(N)$ $\phi^4$ $\beta$-function. In
\cite{17,33,80} the same exponent was used to check the three and four loop
mass anomalous dimension of the $\sigma$ field in (\ref{lag6}). Therefore, the
same reasoning applies here and the expansion of $\omega$ in powers of $1/N$
and $\epsilon$ where $d$~$=$~$8$~$-$~$2\epsilon$ means that it has to be
consistent with the anomalous dimension of $m_3$ in $L_{\phi\, m}^{(8)}$ which
is what we have found. It is possible to carry out a similar analysis for the
cubic operator of $L_{\phi\, m}^{(8)}$ and found
\begin{equation}
\gamma_{m_4}(g_1,g_2,g_3) ~=~ -~ \left[ 9 N g_1^2 + 152 g_2^2 + 180 g_3^2
\right] \frac{1}{720} ~+~ O(g_i^4) ~.
\end{equation}
Extracting the $O(\epsilon)$ term at $O(1/N)$ of the critical anomalous 
dimension we find that it agrees with the exponent derived in \cite{81,82,83}. 
To determine this anomalous dimension we inserted the cubic operator in a 
$\sigma$ $3$-point function at the fully symmetric point. This was to ensure 
the infrared safeness of the ultraviolet renormalization.  

Having established the equivalence of the renormalization group functions with
lower dimensional theories, the next task is to briefly analyse the fixed point
structure. The first issue is to see if there is a conformal window. Again we 
follow \cite{15,17} and solve for the value of $N$ where 
\begin{equation}
\beta_1(g_1,g_2,g_3) ~=~ \beta_2(g_1,g_2,g_3) ~=~ 
\beta_3(g_1,g_2,g_3) ~=~ 0 ~~,~~ 
\det \left( \frac{\partial \beta_i}{\partial g_j} \right) ~=~ 0
\label{confcond}
\end{equation} 
where the first three equations determine the values of the couplings at the
conformal window and the final equation relates to where there are zero
eigenvalues of the $\beta$-function Hessian. In solving these equations we find
several real solutions for $N$ but only three are positive. These are at
$N_{\mbox{\footnotesize{cr}}}$~$=$~$0.006773$, $0.043641$ and $0.1097804$. So 
in effect there is no conformal window unlike the six dimensional case. To give
a flavour of what the fixed point structure looks like at leading order we 
have solved  
\begin{equation}
\beta_1(g_1,g_2,g_3) ~=~ \beta_2(g_1,g_2,g_3) ~=~ 
\beta_3(g_1,g_2,g_3) ~=~ 0 
\label{critcond}
\end{equation} 
for the value $N$~$=$~$500$. This is partly to compare with a similar analysis 
in the gauge theory case. Aside from the trivial solution we found the 
following fixed points, labelled with a subscript,
\begin{eqnarray}
x_{(1)} &=& 0.806810 + 0.262894 i ~+~ O(\epsilon) ~~,~~
y_{(1)} ~=~ - 0.606634 + 11.002012 i ~+~ O(\epsilon) \nonumber \\
z_{(1)} &=& 170.524352 - 1.949764 i ~+~ O(\epsilon) \nonumber \\
x_{(2)} &=& 0.806810 + 0.262894 i ~+~ O(\epsilon) ~~,~~
y_{(2)} ~=~ - 0.606634 + 11.002012 i ~+~ O(\epsilon) \nonumber \\
z_{(2)} &=& 34.132430 + 10.748863 i ~+~ O(\epsilon) \nonumber \\
x_{(3)} &=& - 0.806810 + 0.262894 i ~+~ O(\epsilon) ~~,~~
y_{(3)} ~=~ 0.606634 + 11.002012 i ~+~ O(\epsilon) \nonumber \\
z_{(3)} &=& 34.132430 - 10.748863 i ~+~ O(\epsilon) \nonumber \\
x_{(4)} &=& - 0.806810 - 0.262894 i ~+~ O(\epsilon) ~~,~~
y_{(4)} ~=~ 0.606634 - 11.002012 i ~+~ O(\epsilon) \nonumber \\
z_{(4)} &=& 170.524352 - 1.949764 i ~+~ O(\epsilon) \nonumber \\
x_{(5)} &=& 0.849156 ~+~ O(\epsilon) ~~,~~
y_{(5)} ~=~ - 8.093663 ~+~ O(\epsilon) ~~,~~
z_{(5)} ~=~ - 22.804535 ~+~ O(\epsilon) \nonumber \\
x_{(6)} &=& 0.849156 ~+~ O(\epsilon) ~~,~~
y_{(6)} ~=~ - 8.093663 ~+~ O(\epsilon) ~~,~~
z_{(6)} ~=~ - 97.139996 ~+~ O(\epsilon) \nonumber \\
x_{(7)} &=& O(\epsilon) ~~,~~
y_{(7)} ~=~ 7.694838 ~+~ O(\epsilon) ~~,~~
z_{(7)} ~=~ - 21.929825 ~+~ O(\epsilon) \nonumber \\
x_{(8)} &=& O(\epsilon) ~~,~~
y_{(8)} ~=~ 7.694838 ~+~ O(\epsilon) ~~,~~
z_{(8)} ~=~ - 63.157895 ~+~ O(\epsilon) \nonumber \\
x_{(9)} &=& O(\epsilon) ~~,~~
y_{(9)} ~=~ O(\epsilon) ~~,~~
z_{(9)} ~=~ 16.666667 ~+~ O(\epsilon)
\end{eqnarray}
in the same notation as the large $N$ analysis. In this list we have not 
included simple reflections $g_i$~$\to$~$-$~$g_i$ or complex conjugate partner 
solutions. For those solutions where there are real and imaginary parts for a 
fixed point coupling constant the corresponding critical point anomalous
dimensions are complex. So there are several cases where real anomalous
dimensions for critical $\gamma_\phi(g_1,g_2,g_3)$ and 
$\gamma_\sigma(g_1,g_2,g_3)$ emerge. Only solution $5$ is stable. Using the 
same labelling as for the critical points for the cases where we have real 
exponents we have, for example
\begin{eqnarray}
\left. \gamma_\phi(g_1,g_2,g_3) \right|_{(5)} &=&
-~ 0.014421 \epsilon ~+~ O(\epsilon^2) ~~,~~
\left. \gamma_\sigma(g_1,g_2,g_3) \right|_{(5)} ~=~
0.604609 \epsilon ~+~ O(\epsilon^2) \nonumber \\
\left. \gamma_\phi(g_1,g_2,g_3) \right|_{(6)} &=&
-~ 0.014421 \epsilon ~+~ O(\epsilon^2) ~~,~~
\left. \gamma_\sigma(g_1,g_2,g_3) \right|_{(6)} ~=~
0.604609 \epsilon ~+~ O(\epsilon^2) \nonumber \\
\left. \gamma_\phi(g_1,g_2,g_3) \right|_{(7)} &=& O(\epsilon^2) ~~,~~
\left. \gamma_\sigma(g_1,g_2,g_3) \right|_{(7)} ~=~
-~ 0.105263 \epsilon ~+~ O(\epsilon^2) \nonumber \\
\left. \gamma_\phi(g_1,g_2,g_3) \right|_{(8)} &=& O(\epsilon^2) ~~,~~
\left. \gamma_\sigma(g_1,g_2,g_3) \right|_{(8)} ~=~
-~ 0.105263 \epsilon ~+~ O(\epsilon^2) \nonumber \\
\left. \gamma_\phi(g_1,g_2,g_3) \right|_{(9)} &=& O(\epsilon^2) ~~,~~
\left. \gamma_\sigma(g_1,g_2,g_3) \right|_{(9)} ~=~ O(\epsilon^2) ~.
\end{eqnarray}
Several features emerge, which it transpires will be similar in the gauge
theory case, and that is that different fixed points have the same leading 
order values for the wave function exponents. There is nothing deeply 
significant about this. It is mainly due to absence of $g_3$ in the 
corresponding one loop anomalous dimensions. Where those exponents have the
same critical values the fixed points only differ in the leading order critical
value for $g_3$. The results for fixed points numbered $7$, $8$ and $9$ are
special cases. For these the value for the coupling at criticality means that
$\phi^i$ is in effect a free field. Therefore, the exponents correspond to a
theory which only involves the $\sigma$ field in effect. For instance, solution
$9$ in essence is the eight dimensional single field $\phi^4$ theory when the
propagator has a double pole.

\sect{Eight dimensional $Sp(N)$ scalar theory.}

While the fixed point structure of the $O(N)$ eight dimensional scalar theory
(\ref{lag8}) does not appear as rich as the six dimensional counterpart in that
the conformal window reaches down to small $N$, there is a related scalar
theory which does run parallel to (\ref{lag6}). This is the eight dimensional
version of (\ref{lag8}) but where the symmetry group is $Sp(N)$. Such a 
variation of the scalar theories was considered in six dimensions in 
\cite{40,84}. It involves the presence of an anti-commuting scalar, similar to 
$\phi^i$, which carries the symplectic property. However, it was shown in those
articles that the renormalization group functions could be simply derived from
those of the $O(N)$ model by making the map $N$~$\to$~$-$~$N$. Therefore, if we
repeat this for the renormalization group functions of (\ref{lag8}) we will be 
able to analyse the $Sp(N)$ version. The first step is to ascertain if there is
a conformal window and again we solve (\ref{confcond}) but use
\begin{equation}
\tilde{x} ~=~ i x ~~,~~
\tilde{y} ~=~ i y ~~,~~
\tilde{z} ~=~ -~ z 
\end{equation}
instead. In this instance we find a set of solutions given by
\begin{eqnarray}
N_{(A)} &=& 13563.468614 ~+~ O(\epsilon) ~~,~~
\tilde{x}_{(A)} ~=~ 1.008162 ~+~ O(\epsilon) \nonumber \\
\tilde{y}_{(A)} &=& -~ 16.322777 ~+~ O(\epsilon) ~~,~~
\tilde{z}_{(A)} ~=~ 4.533577 ~+~ O(\epsilon) \nonumber \\
N_{(B)} &=& 6720.118606 ~+~ O(\epsilon) ~~,~~
\tilde{x}_{(B)} ~=~ 1.015639 ~+~ O(\epsilon) \nonumber \\
\tilde{y}_{(B)} &=& -~ 19.355633 ~+~ O(\epsilon) ~~,~~
\tilde{z}_{(B)} ~=~ -~ 202.850049 ~+~ O(\epsilon) \nonumber \\
N_{(C)} &=& 6145.191926 ~+~ O(\epsilon) ~~,~~
\tilde{x}_{(C)} ~=~ 1.014734 ~+~ O(\epsilon) \nonumber \\
\tilde{y}_{(C)} &=& -~ 22.265284 ~+~ O(\epsilon) ~~,~~
\tilde{z}_{(C)} ~=~ -~ 188.134273 ~+~ O(\epsilon) \nonumber \\
N_{(D)} &=& 6145.191926 ~+~ O(\epsilon) ~~,~~
\tilde{x}_{(D)} ~=~ 1.014734 ~+~ O(\epsilon) \nonumber \\
\tilde{y}_{(D)} &=& -~ 22.265284 ~+~ O(\epsilon) ~~,~~
\tilde{z}_{(D)} ~=~ -~ 446.807837 ~+~ O(\epsilon) \nonumber \\
N_{(E)} &=& 2.894045 ~+~ O(\epsilon) ~~,~~
\tilde{x}_{(E)} ~=~ 0.197977 i ~+~ O(\epsilon) \nonumber \\
\tilde{y}_{(E)} &=& -~ 0.456225 i ~+~ O(\epsilon) ~~,~~
\tilde{z}_{(E)} ~=~ 0.215506 ~+~ O(\epsilon) \nonumber \\
N_{(F)} &=& 1.345536 i + 6.030563 ~+~ O(\epsilon) ~~,~~
\tilde{x}_{(F)} ~=~ 0.276745 i + 0.025867 ~+~ O(\epsilon) \nonumber \\
\tilde{y}_{(F)} &=& -~ 0.686337 i + 0.344352 ~+~ O(\epsilon) ~~,~~
\tilde{z}_{(F)} ~=~ 0.383205 i + 0.186459 ~+~ O(\epsilon) 
\end{eqnarray}
where we have omitted the conjugate solution to $F$ to save space. It turns out
that there are several real solutions for the value of $N$ where the number of
real eigenvalues change. These are 
$N_{\mbox{\footnotesize{cr}}}$~$=$~$13564$, $6721$, $6146$, and $3$. 

Given the several ranges for the windows, we have analysed representative 
values of $N$ in order to see the structure of the fixed points for each 
sector by solving (\ref{critcond}). It turns out that the behaviour varies from
sector to sector. Therefore, we provide a set of fixed points for various
representative values of $N$. For instance, when $N$~$=$~$15000$ we have the 
critical couplings 
\begin{eqnarray}
\tilde{x}_{(1),15000} &=& 1.007382 ~+~ O(\epsilon) ~~,~~
\tilde{y}_{(1),15000} ~=~ -~ 16.164156 ~+~ O(\epsilon) \nonumber \\
\tilde{z}_{(1),15000} &=& 103.672328 ~+~ O(\epsilon) \nonumber \\
\tilde{x}_{(2),15000} &=& 1.007382 ~+~ O(\epsilon) ~~,~~
\tilde{y}_{(2),15000} ~=~ -~ 16.164156 ~+~ O(\epsilon) \nonumber \\
\tilde{z}_{(2),15000} &=& -~ 37.868526 ~+~ O(\epsilon) \nonumber \\
\tilde{x}_{(3),15000} &=& 0.974832 ~+~ O(\epsilon) ~~,~~
\tilde{y}_{(3),15000} ~=~ -~ 47.393461 ~+~ O(\epsilon) \nonumber \\
\tilde{z}_{(3),15000} &=& -~ 735.06222 ~+~ O(\epsilon) \nonumber \\
\tilde{x}_{(4),15000} &=& 0.974832 ~+~ O(\epsilon) ~~,~~
\tilde{y}_{(4),15000} ~=~ -~ 47.393461 ~+~ O(\epsilon) \nonumber \\
\tilde{z}_{(4),15000} &=& -~ 2674.674316 ~+~ O(\epsilon) \nonumber \\
\tilde{x}_{(5),15000} &=& 0.865512 ~+~ O(\epsilon) ~~,~~
\tilde{y}_{(5),15000} ~=~ 53.493631 ~+~ O(\epsilon) \nonumber \\
\tilde{z}_{(5),15000} &=& -~ 840.729642 ~+~ O(\epsilon) \nonumber \\
\tilde{x}_{(6),15000} &=& 0.865514 ~+~ O(\epsilon) ~~,~~
\tilde{y}_{(6),15000} ~=~ 53.493631 ~+~ O(\epsilon) \nonumber \\
\tilde{z}_{(6),15000} &=& -~ 3827.814755 ~+~ O(\epsilon) \nonumber \\
\tilde{x}_{(7),15000} &=& O(\epsilon) ~~,~~
\tilde{y}_{(7),15000} ~=~ 42.146362 i ~+~ O(\epsilon) ~~,~~
\tilde{z}_{(7),15000} ~=~ 1894.736842 ~+~ O(\epsilon) \nonumber \\
\tilde{x}_{(8),15000} &=& O(\epsilon) ~~,~~
\tilde{y}_{(8),15000} ~=~ 42.146361 i ~+~ O(\epsilon) ~~,~~
\tilde{z}_{(8),15000} ~=~ 657.894737 ~+~ O(\epsilon) \nonumber \\
\tilde{x}_{(9),15000} &=& O(\epsilon) ~~,~~
\tilde{y}_{(9),15000} ~=~ O(\epsilon) ~~,~~
\tilde{z}_{(9),15000} ~=~ -~ 500.000000 ~.
\end{eqnarray}
In these and subsequent fixed point solutions we omit critical points which are
related by reflections or complex conjugates. Similar features are common with
the $O(N)$ theory with $N$~$=$~$500$ such as solutions $7$, $8$ and $9$ which
correspond to the $\phi^i$-free case. Also there are pairs with the same
$\tilde{x}$ and $\tilde{y}$ values but a different value for $\tilde{z}$. The
main difference is that all solutions are real when $\tilde{x}$~$\neq$~$0$. By
contrast examining the $N$~$=$~$10000$ case we find
\begin{eqnarray}
\tilde{x}_{(1),10000} &=& 1.011031 ~+~ O(\epsilon) ~~,~~
\tilde{y}_{(1),10000} ~=~ -~ 17.015872 ~+~ O(\epsilon) \nonumber \\
\tilde{z}_{(1),10000} &=& -~ 74.728621 + 81.472666 i ~+~ O(\epsilon) 
\nonumber \\
\tilde{x}_{(2),10000} &=& 0.989137 ~+~ O(\epsilon) ~~,~~
\tilde{y}_{(2),10000} ~=~ -~ 36.865152 ~+~ O(\epsilon) \nonumber \\
\tilde{z}_{(2),10000} &=& -~ 454.998667 ~+~ O(\epsilon) \nonumber \\
\tilde{x}_{(3),10000} &=& 0.989137 ~+~ O(\epsilon) ~~,~~
\tilde{y}_{(3),10000} ~=~ -~ 36.865152 ~+~ O(\epsilon) \nonumber \\
\tilde{z}_{(3),10000} &=& -~ 1561.607482 ~+~ O(\epsilon) \nonumber \\
\tilde{x}_{(4),10000} &=& 0.854964 ~+~ O(\epsilon) ~~,~~
\tilde{y}_{(4),10000} ~=~ 43.810992 ~+~ O(\epsilon) \nonumber \\
\tilde{z}_{(4),10000} &=& -~ 558.723323 ~+~ O(\epsilon) \nonumber \\
\tilde{x}_{(5),10000} &=& 0.854964 ~+~ O(\epsilon) ~~,~~
\tilde{y}_{(5),10000} ~=~ 43.810992 ~+~ O(\epsilon) \nonumber \\
\tilde{z}_{(5),10000} &=& -~ 2585.830662 ~+~ O(\epsilon) \nonumber \\
\tilde{x}_{(6),10000} &=& O(\epsilon) ~~,~~
\tilde{y}_{(6),10000} ~=~ 34.412360 i ~+~ O(\epsilon) \nonumber \\
\tilde{z}_{(6),10000} &=& 1263.157895 ~+~ O(\epsilon) \nonumber \\
\tilde{x}_{(7),10000} &=& O(\epsilon) ~~,~~
\tilde{y}_{(7),10000} ~=~ 34.412360 i ~+~ O(\epsilon) ~~,~~
\tilde{z}_{(7),10000} ~=~ 438.596491 ~+~ O(\epsilon) \nonumber \\
\tilde{x}_{(8),10000} &=& O(\epsilon) ~~,~~
\tilde{y}_{(8),10000} ~=~ O(\epsilon) ~~,~~
\tilde{z}_{(8),10000} ~=~ -~ 333.333333 ~+~ O(\epsilon) ~.
\end{eqnarray}
Here there is one fully complex solution. In the next lower window the reality
of all solutions is restored since, for example,  
\begin{eqnarray}
\tilde{x}_{(1),6500} &=& 1.015877 ~+~ O(\epsilon) ~~,~~
\tilde{y}_{(1),6500} ~=~ -~ 19.862247 ~+~ O(\epsilon) \nonumber \\
\tilde{z}_{(1),6500} &=& -~ 174.63918 ~+~ O(\epsilon) \nonumber \\
\tilde{x}_{(2),6500} &=& 1.015877 ~+~ O(\epsilon) ~~,~~
\tilde{y}_{(2),6500} ~=~ -~ 19.862247 ~+~ O(\epsilon) \nonumber \\
\tilde{z}_{(2),6500} &=& -~ 272.79598 ~+~ O(\epsilon) \nonumber \\
\tilde{x}_{(3),6500} &=& 1.009679 ~+~ O(\epsilon) ~~,~~
\tilde{y}_{(3),6500} ~=~ -~ 25.636626 ~+~ O(\epsilon) \nonumber \\
\tilde{z}_{(3),6500} &=& -~ 234.623913 ~+~ O(\epsilon) \nonumber \\
\tilde{x}_{(4),6500} &=& 1.009679 ~+~ O(\epsilon) ~~,~~
\tilde{y}_{(4),6500} ~=~ -~ 25.636626 ~+~ O(\epsilon) \nonumber \\
\tilde{z}_{(4),6500} &=& -~ 669.753355 ~+~ O(\epsilon) \nonumber \\
\tilde{x}_{(5),6500} &=& 0.841616 ~+~ O(\epsilon) ~~,~~
\tilde{y}_{(5),6500} ~=~ 35.419872 ~+~ O(\epsilon) \nonumber \\
\tilde{z}_{(5),6500} &=& -~ 360.906735 ~+~ O(\epsilon) \nonumber \\
\tilde{x}_{(6),6500} &=& 0.841616 ~+~ O(\epsilon) ~~,~~
\tilde{y}_{(6),6500} ~=~ 35.419872 ~+~ O(\epsilon) \nonumber \\
\tilde{z}_{(6),6500} &=& -~ 1704.819491 ~+~ O(\epsilon) \nonumber \\
\tilde{x}_{(7),6500} &=& O(\epsilon) ~~,~~
\tilde{y}_{(7),6500} ~=~ 27.744132 i ~+~ O(\epsilon) ~~,~~
\tilde{z}_{(7),6500} ~=~ 821.052632 ~+~ O(\epsilon) \nonumber \\
\tilde{x}_{(8),6500} &=& O(\epsilon) ~~,~~
\tilde{y}_{(8),6500} ~=~ 27.744132 i ~+~ O(\epsilon) ~~,~~
\tilde{z}_{(8),6500} ~=~ 285.087720 ~+~ O(\epsilon) \nonumber \\
\tilde{x}_{(9),6500} &=& O(\epsilon) ~~,~~
\tilde{y}_{(9),6500} ~=~ O(\epsilon) ~~,~~
\tilde{z}_{(9),6500} ~=~ -~ 216.666667 ~+~ O(\epsilon)
\end{eqnarray}
when $N$~$=$~$6500$. The solutions in this region in effect has the same 
structure as that for $N$~$>$~$13563$. Dropping to the next sector two purely
complex solutions emerge. For instance, when $N$~$=$~$100$ we find  
\begin{eqnarray}
\tilde{x}_{(1),100} &=& 0.504796 + 0.886070 i ~+~ O(\epsilon) ~~,~~
\tilde{y}_{(1),100} ~=~ 1.604579 - 6.710841 i ~+~ O(\epsilon) \nonumber \\
\tilde{z}_{(1),100} &=& 48.95388 + 37.45947 i ~+~ O(\epsilon) \nonumber \\
\tilde{x}_{(2),100} &=& 0.504796 + 0.886070 i ~+~ O(\epsilon) ~~,~~
\tilde{y}_{(2),100} ~=~ 1.604579 - 6.710841 i ~+~ O(\epsilon) \nonumber \\
\tilde{z}_{(2),100} &=& 17.146965 + 5.514596 i ~+~ O(\epsilon) \nonumber \\
\tilde{x}_{(3),100} &=& 0.567011 ~+~ O(\epsilon) ~~,~~
\tilde{y}_{(3),100} ~=~ 3.980936 ~+~ O(\epsilon) \nonumber \\
\tilde{z}_{(3),100} &=& -~ 3.101886 ~+~ O(\epsilon) \nonumber \\
\tilde{x}_{(4),100} &=& 0.567011 ~+~ O(\epsilon) ~~,~~
\tilde{y}_{(4),100} ~=~ 3.980936 ~+~ O(\epsilon) \nonumber \\
\tilde{z}_{(4),100} &=& -~ 25.322927 ~+~ O(\epsilon) \nonumber \\
\tilde{x}_{(5),100} &=& O(\epsilon) ~~,~~
\tilde{y}_{(5),100} ~=~ 3.441236 i ~+~ O(\epsilon) ~~,~~
\tilde{z}_{(5),100} ~=~ 12.631579 ~+~ O(\epsilon) \nonumber \\
\tilde{x}_{(6),100} &=& O(\epsilon) ~~,~~
\tilde{y}_{(6),100} ~=~ 3.441236 i ~+~ O(\epsilon) ~~,~~
\tilde{z}_{(6),100} ~=~ 4.385965 ~+~ O(\epsilon) \nonumber \\
\tilde{x}_{(7),100} &=& O(\epsilon) ~~,~~
\tilde{y}_{(7),100} ~=~ O(\epsilon) ~~,~~
\tilde{z}_{(7),100} ~=~ -~ 3.333333 ~+~ O(\epsilon) ~.
\end{eqnarray}
Throughout each of these solutions one of the real fixed points is the one 
which the large $N$ exponents in the $Sp(N)$ version of (\ref{lag8}) are
connected to.

One final example is of special interest. When $N$~$=$~$2$ our solutions to
(\ref{critcond}) are
\begin{eqnarray}
x_{(1),2} &=& 0.193438 i ~+~ O(\epsilon) ~~,~~
y_{(1),2} ~=~ -~ 0.269684 i ~+~ O(\epsilon) \nonumber \\
z_{(1),2} &=& 0.091641 ~+~ O(\epsilon) \nonumber \\
x_{(2),2} &=& 0.193438 i ~+~ O(\epsilon) ~~,~~
y_{(2),2} ~=~ -~ 0.269684 i ~+~ O(\epsilon) \nonumber \\
z_{(2),2} &=& -~ 0.038310 ~+~ O(\epsilon) \nonumber \\
x_{(3),2} &=& 0.149071 i ~+~ O(\epsilon) ~~,~~
y_{(3),2} ~=~ -~ 0.447214 i ~+~ O(\epsilon) \nonumber \\
z_{(3),2} &=& 0.207407 ~+~ O(\epsilon) \nonumber \\
x_{(4),2} &=& 0.149071 i ~+~ O(\epsilon) ~~,~~
y_{(4),2} ~=~ -~ 0.447214 i ~+~ O(\epsilon) \nonumber \\
z_{(4),2} &=& 0.066667 ~+~ O(\epsilon) \nonumber \\
x_{(5),2} &=& 0.282351 ~+~ O(\epsilon) ~~,~~
y_{(5),2} ~=~ 0.433979 ~+~ O(\epsilon) ~~,~~
z_{(5),2} ~=~ 0.027985 ~+~ O(\epsilon) \nonumber \\
x_{(6),2} &=& 0.282351 ~+~ O(\epsilon) ~~,~~
y_{(6),2} ~=~ 0.433979 ~+~ O(\epsilon) ~~,~~
z_{(6),2} ~=~ -~ 0.407685 ~+~ O(\epsilon) \nonumber \\
x_{(7),2} &=& O(\epsilon) ~~,~~
y_{(7),2} ~=~ 0.486664 i ~+~ O(\epsilon) ~~,~~
z_{(7),2} ~=~ 0.252632 ~+~ O(\epsilon) \nonumber \\
x_{(8),2} &=& O(\epsilon) ~~,~~
y_{(8),2} ~=~ 0.486664 i ~+~ O(\epsilon) ~~,~~
z_{(8),2} ~=~ 0.087719 ~+~ O(\epsilon) \nonumber \\
x_{(9),2} &=& O(\epsilon) ~~,~~
y_{(9),2} ~=~ O(\epsilon) ~~,~~
z_{(9),2} ~=~ -~ 0.066667 ~+~ O(\epsilon) ~.
\end{eqnarray}
While there are fewer purely real solutions those that are imaginary only for
$\tilde{x}$ and $\tilde{y}$ will have real squares when put on the same
footing as $\tilde{z}$. The main observation is solution $4$, which is a stable
fixed point, has the property that 
\begin{equation}
\tilde{y} ~=~ 3 \tilde{x} ~+~ O(\epsilon) ~~,~~ 
\tilde{z} ~=~ 3 \tilde{x}^2 ~+~ O(\epsilon) ~.
\end{equation}
This is not an accident as a similar solution emerged in the six dimensional
$Sp(N)$ case for $N$~$=$~$2$, \cite{84}, although there was no quartic 
interaction there. In \cite{84} it was shown to be due to a hidden 
supersymmetry based on the supergroup $OSp(1|2)$. Thus it would appear that the
same symmetry arises in the eight dimensional scalar theory. One property of 
this supersymmetry is that the field anomalous dimension for $\phi^i$ and 
$\sigma$ should be equivalent and we have checked this and found that 
\begin{equation}
\left. \gamma_\phi(g_1,g_2,g_3) \right|_{(4),2} ~=~
\left. \gamma_\sigma(g_1,g_2,g_3) \right|_{(4),2} ~=~ -~ 0.111111 \epsilon ~+~ 
O(\epsilon^2) ~. 
\end{equation}
Actually the same leading order exponents emerge for solution $3$ too but this
is only due to $\gamma_\phi(g_1,g_2,g_3)$ and $\gamma_\sigma(g_1,g_2,g_3)$ not
depending on $g_3$ at one loop. What is perhaps more intriguing is that the
critical point structure of the six dimensional $Sp(2)$ case is given by the
$q$~$\to$~$0$ limit of the $q$-state Potts model, \cite{85}. In \cite{86} it 
was suggested that the upper critical dimension for this equivalence was six. 
Given the relation of (\ref{lag8}) now with (\ref{lag6}) at the Wilson-Fisher 
fixed point and the appearance of a hidden symmetry for $Sp(2)$ at a specific 
fixed point, similar to six dimensions \cite{84}, it would be interesting to 
see whether the restriction to six dimensions argued in \cite{86} could be 
extended to eight dimensions. While our focus in this section has been on the 
$Sp(N)$ theory, which reveals a rich fixed point spectrum on a par with that of
(\ref{lag6}), \cite{17}, a full analysis would require higher order 
computations.

\sect{Higher dimensional gauge theories.}

Having discussed a model scalar theory of some of the structural similarities
to higher dimensional gauge theories we now concentrate on the construction of
the six dimensional QCD Lagrangian in this section. In essence the core 
properties of eight dimensional $O(N)$ $\phi^3$ theory translate to the QCD 
case. The main difference is the presence of gauge symmetry which requires a 
modification of our algorithm for the completion of the higher dimensional 
theory and the construction of the tower of theories which are equivalent at 
the Wilson-Fisher fixed point. First, we recall the four dimensional QCD 
Lagrangian is 
\begin{equation}
L^{(4)} ~=~ -~ \frac{1}{4} G_{\mu\nu}^a
G^{a \, \mu\nu} ~-~ \frac{1}{2\alpha} (\partial^\mu A^a_\mu)^2 ~-~
\bar{c}^a \left( \partial^\mu D_\mu c \right)^a ~+~ 
i \bar{\psi}^{iI} \Dslash \psi^{iI}
\label{lagqcd4}
\end{equation}
where $A^a_\mu$ is the gluon, $\psi^{iI}$ is the quark and $c^a$ are the 
Faddeev-Popov ghost fields. Here the indices take the ranges
$1$~$\leq$~$a$~$\leq$~$\NA$, $1$~$\leq$~$I$~$\leq$~$\NF$ and
$1$~$\leq$~$i$~$\leq$~$\Nf$ where $\NF$ and $\NA$ are the respective dimensions
of the fundamental and adjoint representations of the colour group and $\Nf$ is
the number of quark flavours. Also $D_\mu$ is the covariant derivative
and $G^a_{\mu\nu}$ is the field strength. Throughout we choose to work with the
canonical linear covariant gauge fixing whose associated gauge parameter is 
$\alpha$. While this is a standard Lagrangian it is worth noting several 
features relevant to the present discussion. The Lagrangian is constructed in 
several stages. The first is to write down all independent local gauge 
invariant operators which are built from the $A^a_\mu$ and $\psi^{iI}$ fields 
and are renormalizable in the dimension of interest which is four for 
(\ref{lagqcd4}). For the moment we will exclude lower dimensional operators 
which would introduce masses. Unlike scalar theories such gauge invariant 
Lagrangians produce fields with more degrees of freedom than are present in 
nature and therefore a gauge fixing is required. Again this gauge fixing, which
does not have to be covariant or linear as we are choosing here, has to be 
local, renormalizable and of dimension four. The gauge fixing terms 
subsequently breaks gauge invariance. So one instead requires that the
Lagrangian is BRST invariant rather than gauge invariant. These considerations
clearly have been satisfied in (\ref{lagqcd4}).  

One ingredient from our earlier algorithm appears to have been omitted in this
instance and that is the theory in two dimensions with the same symmetries
which is connected via the Wilson-Fisher fixed point. In other words the base 
theory which is in the same universality class. This requires some care given 
the nature of a two dimensional spin-$1$ field. It transpires that the 
equivalent theory is the non-abelian Thirring model (NATM), \cite{60}, which
has the Lagrangian 
\begin{equation}
L^{(2)} ~=~ i \bar{\psi}^{iI} \partialslash \psi^{iI} ~+~ 
\frac{g}{2} \left( \bar{\psi}^{iI} T^a_{IJ} \gamma^\mu \psi^{iJ} \right)^2
\label{lagnatm}
\end{equation}
where $T^a$ are the colour group generators. Unlike (\ref{lag2}) there can be
no base Lagrangian which is linear in a spin-$1$ without breaking colour and
Lorentz symmetry. As presented the connection with (\ref{lagqcd4}) appears 
distant due to the absence of a field $A^a_\mu$. However, the interaction of 
(\ref{lagnatm}) may be rewritten in two dimensions in terms of an auxiliary 
spin-$1$ field to produce
\begin{equation}
L^{(2)} ~=~ i \bar{\psi}^{iI} \partialslash \psi^{iI} ~+~ 
g \bar{\psi}^{iI} T^a_{IJ} \gamma^\mu \psi^{iJ} A^a_\mu ~-~ \frac{g}{2}
A^a_\mu A^{a\,\mu} ~.
\label{lagqcd2}
\end{equation}
As it stands this version of $L^{(2)}$ appears to be an improvement on 
(\ref{lagnatm}) with regard to equivalence with (\ref{lagqcd4}) but it does not
appear to be consistent with our completion argument. One objection is the
apparent absence of gauge invariance and by association the gauge fixing and
ghost terms. On the contrary the equivalence with four dimensional QCD 
observed in \cite{60} has subsequently been verified computationally to several
orders in the large $\Nf$ expansion in \cite{61,62,63}. The bridge is in the 
main two fold. First, the auxiliary field reformulation is a strictly two 
dimensional relation. Second, to proceed with the large $\Nf$ analysis through 
the connecting Wilson-Fisher fixed point the key is the quark-gluon vertex 
which together with the quark kinetic term define the canonical dimensions of 
the field in the $d$-dimensional universality class. The sector which is purely
gluonic, such as $\half A^a_\mu A^{a\,\mu}$ in (\ref{lagqcd2}) and 
$G^a_{\mu\nu} G^{a\,\mu\nu}$ in (\ref{lagqcd4}), in essence define the 
canonical dimensions of the respective coupling constants in each theory. Of 
course, the couplings have different dimensionalities in renormalizable 
theories in different spacetime dimensions. Therefore, in the
large $\Nf$ approach discussed in \cite{63}, the gauge fixed Lagrangian at 
criticality has an analytically regularized gauge fixing with associated
Faddeev-Popov ghost sector modifications, \cite{63}. In other words formally
\begin{eqnarray}
L^{\mbox{\footnotesize{NATM}}} &=&
i \bar{\psi}^{iI} \partialslash \psi^{iI} ~+~ 
g \bar{\psi}^{iI} T^a_{IJ} \gamma^\mu \psi^{iJ} A^a_\mu ~-~
\bar{c}^a \left( \partial^\mu D_\mu c \right)^a \nonumber \\
&& -~ \frac{g}{2} A^a_\mu A^{a\,\mu} ~+~
\frac{1}{2\alpha} \left( \partial^\mu A^a_\mu \right) \frac{1}{\Box^{4-d}}
\left( \partial^\nu A^a_\nu \right) 
\label{gfd}
\end{eqnarray}
is used to determine the large $\Nf$ critical exponents, \cite{63}. One 
immediate objection to this is that one does not have locality. Equally one 
also loses perturbative renormalizability for the lower dimensional theory but 
at criticality these are not an issue. What one has to accept is that the 
critical equivalence is valid in the Landau gauge which corresponds to 
$\alpha$~$=$~$0$. This is more subtle than it appears and is not unrelated to 
our algorithm extended to the gauge theory context. In two dimensions we have 
treated $A^a_\mu$ as an auxiliary spin-$1$ field. If it were a gauge field in 
the context of (\ref{lagqcd4}) then clearly the operator 
$\half A^a_\mu A^{a\,\mu}$ is not gauge invariant. However, it is possible to 
write down several gauge invariant dimension two operators but in this instance
the locality assumption has to be dropped. 

For example, the operator
\begin{equation}
{\cal O}_2 ~=~ -~ \frac{1}{2} G^{a \, \mu\nu} \frac{1}{D^2} G^{a \, \mu\nu}
\label{op2}
\end{equation}
is dimension two and gauge invariant but clearly nonlocal. Such an operator
has appeared before, \cite{87,88}, in the context of three dimensional gauge
theories and studied for their relation to temperature QCD. Despite the 
presence of the nonlocality it is possible to localize the operator and 
determine its renormalization to several loop orders, \cite{89,90}. In other 
words this nonlocal operator can be regarded as being perturbatively 
renormalizable. For instance, the one loop anomalous dimension in four 
dimensions is proportional to the one loop QCD $\beta$-function, \cite{89,90}. 
Beyond one loop this proportionality ceases. This is due to the presence of 
extra or ghost fields which arise in the localizing procedure and their 
coupling constants appear in the two loop and higher operator anomalous 
dimension. Another gauge invariant gluonic dimension two operator is
\begin{equation}
{\cal O} ~ \equiv ~ \frac{1}{2} \stackrel{\mbox{\begin{small}min\end{small}}}
{\mbox{\begin{tiny}$\{U\}$\end{tiny}}} \int d^4x \, \left( A^{a \, U}_\mu
\right)^2 
\end{equation}
where $A^{a \, U}_\mu$ is the transport of a gauge field along a gauge orbit
\begin{equation}
A^U_\mu ~=~ U A_\mu U^\dagger ~-~ \frac{i}{g} \left( \partial_\mu U \right)
U^\dagger
\end{equation}
and $U$ is gauge group element. By construction ${\cal O}$ is gauge invariant
and forms the basis for a gauge fixing, \cite{91,92,93,94}, which does not 
suffer from Gribov copy issues. There are various ways of writing ${\cal O}$ 
perturbatively in terms of other nonlocal operators, \cite{91,95}. A gauge 
invariant expansion was given in \cite{91,95} and ${\cal O}_2$ is in fact the 
first term. The three leg operator was presented in \cite{91} and has 
structural similarities to the dimension six operator considered later. More 
recently an algorithm to produce the subsequent operators was given in 
\cite{95}. Despite the nonlocality the one loop renormalization of ${\cal O}$ 
was given in \cite{96}. There it was shown that the gauge parameter was indeed 
absent in the anomalous dimension. While such operators address the issue of 
constructing a gauge invariant dimension two operator, which is present in 
theories connected at the Wilson-Fisher fixed point, there is a connection with
(\ref{lagqcd2}). Although locality is sacrificed for gauge invariance to 
produce a nonlocal operator, both of the operators ${\cal O}_2$ and ${\cal O}$ 
truncate to $\half A^a_\mu A^{a\,\mu}$ when one specifies the Landau gauge. In 
this gauge this operator is also BRST invariant as the ghost mass term is 
absent. The upshot is that as discussed in \cite{63} when comparing our 
perturbative results at the Wilson-Fisher fixed point for gauge theories in the
different dimensions, we can only compare critical exponents which derive from 
gauge dependent renormalization group functions in the Landau gauge. For 
exponents based on gauge independent renormalization group functions this point
will not be relevant.

Returning to the problem of constructing a six dimensional gauge theory the
first stage is to write down the set of independent gauge invariant dimension
six operators with which to build a Lagrangian. For the quark sector to
maintain connectivity with the four dimensional gauge theory the set includes
$i \bar{\psi}^{iI} \Dslash \psi^{iI}$. In six dimensions this immediately
defines the canonical dimension of the quark field to be $\frac{5}{2}$. Thus
unlike two dimensions there are no quartic or higher operators which include 
quark fields. As such an operator would require an anti-quark to ensure a 
Lorentz scalar term one sees that there is only one dimension six quark
operator. This is important since, for instance, when considering six 
dimensional operators in four dimensional QCD effective theories, $4$-fermi 
operators are included in the same discussion. In the six dimensional case they
will not appear in a Lagrangian since such $4$-fermi operators actually have a 
canonical dimension of $10$ and so are absent in a renormalizable Lagrangian. 
Such $4$-fermi operators are only perturbatively renormalizable in two 
dimensions as is evident in (\ref{lagnatm}) or (\ref{lagqcd2}). We now change 
our focus to the gluonic sector. In \cite{41,66} such dimension six gluonic 
operators were considered and it transpires that there are four potential 
candidates which are
\begin{eqnarray}
{\cal O}^{(6)}_1 &=& \left( D_\mu G_{\nu\sigma}^a \right) 
\left( D^\mu G^{a \, \nu\sigma} \right) ~~~,~~~
{\cal O}^{(6)}_2 ~=~ \left( D^\mu G_{\mu\sigma}^a \right) 
\left( D_\nu G^{a \, \nu\sigma} \right) \nonumber \\ 
{\cal O}^{(6)}_3 &=& \left( D_\mu G_{\nu\sigma}^a \right) 
\left( D^\sigma G^{a \, \mu\nu} \right) ~~~,~~~
{\cal O}^{(6)}_4 ~=~ f^{abc} G_{\mu\nu}^a \, G^{b \, \mu\sigma} \, 
G^{c \,\nu}_{~~\,\sigma} ~.
\label{dim6op}
\end{eqnarray}
However, these are not all independent due to either integration by parts or
use of the Bianchi identity
\begin{equation}
D_\mu G^a_{\nu\sigma} ~+~ D_\nu G^a_{\sigma\mu} ~+~ 
D_\sigma G^a_{\mu\nu} ~=~ 0 ~.
\end{equation}
Total derivative operators can be ignored in the Lagrangian construction due to
conservation of energy-momentum. So of the set (\ref{dim6op}) we are free to
choose any two for our six dimensional QCD Lagrangian $L^{(6)}$. In \cite{41} 
${\cal O}^{(6)}_1$ and ${\cal O}^{(6)}_2$ were chosen as the two independent 
operators but we will take a different basis which is ${\cal O}^{(6)}_1$ and 
${\cal O}^{(6)}_4$. The reason for this choice rests partly in the potential 
connection with four dimensions as noted earlier. Thus
if there are fixed points in the six dimensional gauge theory which connect
with the infrared structure of QCD in four dimensions after some sort of
summation, it seems appropriate to include the key operator explicitly with
its own coupling constant at the outset. Moreover, ${\cal O}^{(6)}_1$ is the
natural extension of the gluon kinetic term which is why that is chosen for the
other independent operator. Irrespective of which basis choice we make the 
gluon propagator will now have a double pole. However, if there is connectivity
with the infrared structure of a lower dimensional gauge theory a double pole 
propagator may not be inappropriate. Another reason for taking 
${\cal O}^{(6)}_1$ and ${\cal O}^{(6)}_4$ rests in the nature of the coupling 
constants. If one chose ${\cal O}^{(6)}_2$ instead of ${\cal O}^{(6)}_4$ then 
there is the problem of what relative weight to assign each term. The 
appropriate way to proceed is to introduce a weighting parameter such as 
$\beta$ and include
\begin{equation}
\beta {\cal O}^{(6)}_1 ~+~ (1-\beta) {\cal O}^{(6)}_2
\end{equation}
as the two independent operators in the Lagrangian. The parameter $\beta$ would
not be present in the gluon propagator but would be present in the interaction
terms. It is not a gauge fixing parameter but rather represents a measure of
the interpolation. Thus its renormalization would be independent of the gauge
parameter in $\MSbar$ for instance. In effect in the interaction terms the
product of $\beta$ with $g_1$ corresponds to a second coupling constant which
is independent of $g_1$ and if one were to use this set of operators in the
Lagrangian then $\beta g_1$ would be redefined as a second coupling. While this
is perfectly viable as a strategy it seems more appropriate to use one $2$-leg
operator for the kinetic term and have the second independent operator as 
higher leg which is why we choose ${\cal O}^{(6)}_1$ and ${\cal O}^{(6)}_4$. 
Equally no intermediate interpolating parameter needs to be introduced as one
just couples the latter operator with the independent coupling retaining the
gauge coupling, $g_1$, in the gluon kinetic term. Thus the gauge invariant six 
dimensional Lagrangian, $L^{(6)}_{\mbox{\footnotesize{GI}}}$, of QCD we begin 
with is 
\begin{equation}
L^{(6)}_{\mbox{\footnotesize{GI}}} ~=~ 
-~ \frac{1}{4} \left( D_\mu G_{\nu\sigma}^a \right) 
\left( D^\mu G^{a \, \nu\sigma} \right) ~+~ 
\frac{g_2}{6} f^{abc} G_{\mu\nu}^a \, G^{b \, \mu\sigma} \, 
G^{c \,\nu}_{~~\,\sigma} ~+~ i \bar{\psi}^{iI} \Dslash \psi^{iI} ~.
\label{lagqcd6gi}
\end{equation}
As we have an interaction over and above those which derive from terms
involving the covariant derivative we need to be clear about the notation.
Throughout when additional operators are appended to a gauge theory in higher
dimensions such as here then we will use the coupling constant $g_1$ as that
which appears in the covariant derivative, $D_\mu$, and hence also
$G^a_{\mu\nu}$. For theories with extra symmetries such as supersymmetry the
second coupling, $g_2$, could be related to $g_1$. Equally if one proceeded
with the choice involving $\beta$ its value would be fixed by the extra
symmetry. As an aside effective Lagrangians similar to (\ref{lagqcd6gi}) have
been studied in four dimensions in various covariant and non-covariant gauges 
in order to explore possible non-perturbative behaviour of the gluon propagator
in the infrared region, \cite{97,98,99}. 

The final aspect of our discussion centres on the form of the gauge fixing
terms which need to be present in order to carry out perturbative calculations.
As in four dimensions we choose to fix in an arbitrary linear covariant gauge 
$\partial^\mu A^a_\mu$~$=$~$0$. However, the usual four dimensional gauge 
fixing term in addition to the Faddeev-Popov ghost term which implements this 
condition cannot be used in six dimensions due to the fact that the canonical 
terms are dimension four. Instead motivated by (\ref{gfd}) we use a BRST 
invariant dimension six gauge fixing where the shortfall in dimensionality of 
the operators are made up for by spacetime derivatives. In other words our 
gauge fixed six dimensional QCD Lagrangian is 
\begin{eqnarray}
L^{(6)} &=& -~ \frac{1}{4} \left( D_\mu G_{\nu\sigma}^a \right) 
\left( D^\mu G^{a \, \nu\sigma} \right) ~+~ 
\frac{g_2}{6} f^{abc} G_{\mu\nu}^a \, G^{b \, \mu\sigma} \, 
G^{c \,\nu}_{~~\,\sigma} \nonumber \\
&& -~ \frac{1}{2\alpha} \left( \partial_\mu \partial^\nu A^a_\nu \right)
\left( \partial^\mu \partial^\sigma A^a_\sigma \right) ~-~
\bar{c}^a \Box \left( \partial^\mu D_\mu c \right)^a ~+~ 
i \bar{\psi}^{iI} \Dslash \psi^{iI}
\label{lagqcd6}
\end{eqnarray}
where $\alpha$ is the covariant fixing parameter with the Landau gauge
corresponding to $\alpha$~$=$~$0$. It is straightforward to check that the
Lagrangian is BRST invariant without modification of the canonical BRST
transformations on the fields. The gauge fixing term allows one to find the
gluon propagator since when $\alpha$~$\neq$~$0$ the quadratic part of the
momentum space Lagrangian is invertible. The gluon and ghost propagators are
then 
\begin{eqnarray}
\langle A^a_\mu(p) A^b_\nu(-p) \rangle &=& -~ 
\frac{\delta^{ab}}{(p^2)^2} \left[ \eta_{\mu\nu} ~-~
(1 - \alpha) \frac{p_\mu p_\nu}{p^2} \right] \nonumber \\
\langle c^a(p) \bar{c}^b(-p) \rangle &=& -~ \frac{\delta^{ab}}{(p^2)^2} 
\end{eqnarray}
with the double pole propagator emerging as noted earlier and similar to
\cite{97,98,99}.

We close this section by considering the extension of the Lagrangians to lower
dimensional operators and hence mass terms. This is similar to the scalar
theory case but with the constraint that additional terms have to be gauge
invariant in the first instance and when the gauge is fixed they have to be
BRST invariant. For a gauge theory in $D$-dimensions where $D$ is an integer
the extra operators are no more than $(D-2)$-dimensional for the gluon and
ghost sector. The upshot of this is that the structure is available from the 
lower dimensional Lagrangians discussed above but with the caveat that a quark 
mass operator can be included. This will be common to all gauge theories and is
$(D-1)$-dimensional. We will always denote the quark mass as $m_1$. In four
dimensions there is therefore only one dimension two gluonic operator to be
added in to $L^{(4)}$. If one requires it to be gauge invariant then one has to
use ${\cal O}$ but weaken the locality assumption. Otherwise the only operator
possible is the local BRST mass operator, \cite{100}, in the massive extension 
of (\ref{lagqcd4}) which is
\begin{eqnarray}
L_m^{(4)} &=& L^{(4)} + m_1 \bar{\psi}^{iI} \psi^{iI} ~-~ 
\frac{1}{2} m_2^2 A^a_\mu A^{a \, \mu} ~+~ m_2^2 \alpha \bar{c}^a c^a ~.
\end{eqnarray}
The pattern for six dimensions is straightforward to see and we find that the
extension to (\ref{lagqcd6}) is
\begin{eqnarray}
L_m^{(6)} &=& L^{(6)} + m_1 \bar{\psi}^{iI} \psi^{iI} ~-~ 
\frac{1}{4} m_2^2 G_{\mu\nu}^a G^{a \, \mu\nu} ~-~ 
\frac{1}{2\alpha} m_3^2 (\partial^\mu A^a_\mu)^2 ~-~
m_3^2 \bar{c}^a \left( \partial^\mu D_\mu c \right)^a \nonumber \\
&& -~ \frac{1}{2} m_4^4 A^a_\mu A^{a \, \mu} ~+~ m_4^4 \alpha \bar{c}^a c^a ~. 
\label{lagqcd6m}
\end{eqnarray}
In effect each gauge or BRST invariant lower dimensional operator gains a
separate mass. In essence this is the coupling constant of the corresponding
operator in the lower dimensional theory and across the different dimensions
these operators range from being relevant to irrelevant. While $L_m^{(4)}$ can
only be extended by a BRST invariant operator, in $L_m^{(6)}$ before gauge
fixing one can have a mass associated with a gauge invariant gluonic operator.
To see the effect of such a term it is instructive to derive the propagators
for $L_m^{(6)}$. We have 
\begin{eqnarray}
\langle A^a_\mu(p) A^b_\nu(-p) \rangle &=& -~ 
\frac{\delta^{ab}}{[(p^2)^2 + m_2^2 p^2 + m_4^4 ]} \left[ \eta_{\mu\nu} ~-~
\frac{[ p^2 + m_3^2 - \alpha ( p^2 + m_2^2 )] p_\mu p_\nu}
{[(p^2)^2 + m_3^2 p^2 + \alpha m_4^4 ]} \right] \nonumber \\
\langle c^a(p) \bar{c}^b(-p) \rangle &=& -~ 
\frac{\delta^{ab}}{[(p^2)^2 + m_3^2 p^2 + \alpha m_4^4 ]} 
\end{eqnarray}
for arbitrary $\alpha$. Alternatively one can express the gluon propagator in 
terms of the respective transverse and longitudinal tensors as
\begin{equation}
\langle A^a_\mu(p) A^b_\nu(-p) \rangle ~=~ -~
\delta^{ab} \left[ \frac{P_{\mu\nu}(p)}{[(p^2)^2 + m_2^2 p^2 + m_4^4 ]} ~+~ 
\frac{\alpha L_{\mu\nu}(p)}{[(p^2)^2 + m_3^2 p^2 + \alpha m_4^4 ]} \right]
\end{equation}
where
\begin{equation}
P_{\mu\nu}(p) ~=~ \eta_{\mu\nu} ~-~ \frac{p_\mu p_\nu}{p^2} ~~,~~
L_{\mu\nu}(p) ~=~ \frac{p_\mu p_\nu}{p^2} ~. 
\end{equation}
In this formulation the connection of the longitudinal part of the gluon 
propagator with the ghost propagator is clearer. As an aside the gluon 
propagator takes a simpler form in the Feynman gauge $\alpha$~$=$~$1$. If in
addition, for instance, it were the case that $m_2$~$=$~$m_3$ then the gluon
propagator would simplify further and only involve $\eta_{\mu\nu}$ similar to
the completely massless theory for this specific gauge. However, this mass 
equality condition would require an additional symmetry in order to have this 
simplification. In the case when there is only a gauge invariant dimension four
mass operator the propagators reduce to 
\begin{eqnarray}
\left. \langle A^a_\mu(p) A^b_\nu(-p) \rangle \right|_{m_3=m_4=0} &=& 
-~ \frac{\delta^{ab}}{p^2[p^2+m_2^2]} \left[ \eta_{\mu\nu} ~-~
\frac{p_\mu p_\nu}{p^2} \right] ~-~ 
\alpha \delta^{ab} \frac{p_\mu p_\nu}{(p^2)^3} \nonumber \\
\left. \langle c^a(p) \bar{c}^b(-p) \rangle \right|_{m_3=m_4=0} &=& 
-~ \frac{\delta^{ab}}{(p^2)^2} 
\end{eqnarray}
so that this mass operator removes the double pole propagator. The double pole
remains in the ghost propagator to account for the corresponding pole in the
longitudinal part of the gluon propagator. Another limit to consider is that of
the Landau gauge as it will transpire that $\alpha$~$=$~$0$ is a fixed point of
the renormalization group flow. Then we have 
\begin{eqnarray}
\left. \langle A^a_\mu(p) A^b_\nu(-p) \rangle \right|_{\alpha=0} &=& -~ 
\frac{\delta^{ab}}{[(p^2)^2 + m_2^2 p^2 + m_4^4 ]} \left[ \eta_{\mu\nu} ~-~
\frac{p_\mu p_\nu}{p^2} \right] \nonumber \\
\left. \langle c^a(p) \bar{c}^b(-p) \rangle \right|_{\alpha=0} &=& -~ 
\frac{\delta^{ab}}{p^2[p^2 + m_3^2 ]} 
\label{prop6m}
\end{eqnarray}
so that the gluon propagator has a denominator similar to that of a Stingl 
propagator, \cite{79}. The form of these massive Landau gauge propagators is 
interesting in respect of the current understanding of the infrared behaviour 
of the four dimensional gluon propagator. Briefly, lattice analyses of the 
gluon and Faddeev-Popov ghost propagators in the zero momentum limit indicate 
that the gluon propagator freezes to a non-zero finite value while the ghost 
propagator behaves like $1/p^2$. This has been observed in a variety of 
non-perturbative studies. For instance, the present situation can be found in a
representative set of articles, \cite{49,50,51,52,53,54,55,56,57,58,59}. This 
low energy behaviour has been modelled directly in four dimensions with various
approaches including a modification of the Gribov Lagrangian, \cite{89,90,101}.
It would be interesting to see if the lattice data could be modelled with the
parametrization of (\ref{prop6m}). This would require appending a numerator
parameter for each propagator. However, if the infrared behaviour derives from 
a non-perturbative fixed point in four dimensional QCD accessing it in 
perturbation theory will not be viable. On the contrary if a fixed point in six
dimensional QCD is in the same universality class as this infrared one in four 
dimensions then it may be the case that it will be computationally accessible 
from the higher dimensional theory. Though it would require high loop 
calculations and summation methods to quantify the qualitative behaviour we 
have presented. Intriguingly the Schwinger-Dyson analysis of \cite{98} produced
an effective infrared QCD Lagrangian in four dimensions whose gauge invariant 
part involved the two gluonic operators of (\ref{lagqcd6gi}) together with a 
mass scale necessary to balance the dimensionality. In some sense this gives 
weight to the idea that a perturbatively accessible fixed point of the actual 
six dimensional Lagrangian of (\ref{lagqcd6gi}) could be in same universality 
class of an infrared or non-perturbative fixed point in four dimensional QCD. 
While lattice evidence, \cite{49,50,51,52,53,54,55,56,57,58,59}, in recent 
years suggests a non-scaling gluon propagator in the low momentum region, the 
additional freedom provided by lower dimensional operators in (\ref{lagqcd6m}),
which appear to give propagators qualitatively consistent with the data, could 
be regarded as corrections to the scaling behaviour in the neighbourhood of the
fixed point. What is also apparent is the parallel relation of (\ref{lag6}) and
(\ref{lag8}). A toy $\phi^3$ theory was examined in \cite{67,68} as a model of 
QCD but equally the eight dimensional partner has propagator structures 
parallel to the infrared gluon propagator behaviour in \cite{98}. Finally in 
comparing (\ref{prop6m}) with the corresponding form in the models of 
\cite{101} it is interesting to contrast the nature of the operators which 
correspond to the masses of the gluon propagator. In \cite{101} $m_2$ coupled 
to the dimension two BRST invariant gluon mass operator, which is local in the 
Landau gauge, while $m_4$ was associated with the Landau gauge Gribov operator 
which is nonlocal and dimension zero.  

\sect{Large $\Nf$ expansion.}

As we will be using large $\Nf$ results to compare our higher dimensional
perturbative QCD results it is worth relating relevant aspects to our 
Lagrangian construction. It is based on the observation of \cite{60} that QCD 
and the NATM are in the same universality class. In other words the connecting 
interaction is the quark-gluon vertex but for the $d$-dimensional critical 
point large $\Nf$ construction of \cite{35,36} one has to reformulate 
(\ref{lagqcd2}) in a slightly different way at the outset. Beginning from 
(\ref{lagnatm}) we rewrite it as
\begin{equation}
L^{(2)} ~=~ i \bar{\psi}^{iI} \partialslash \psi^{iI} ~+~
\bar{\psi}^{iI} \gamma^\mu T^a_{IJ} \psi^{iJ} \tilde{A}^a_\mu ~-~ \frac{1}{2g}
\tilde{A}^a_\mu \tilde{A}^{a \, \mu}
\label{largenf2}
\end{equation}
at criticality in preparation for large $\Nf$. The main reason why the coupling
constant has been rescaled into the spin-$1$ field is that the interaction is 
common to all theories in the universality class. The coupling constants have 
different dimensions and are themselves not universal being tied to each theory
in the integer dimensions. In other words they are the couplings of different 
operators in the overall universal theory but their associated operator is only
relevant in the critical sense in a particular spacetime dimension. A similar 
rescaling in $L^{(4)}$ would produce the same interaction as (\ref{largenf2}) 
but with the new coupling appearing in front of the 
$G_{\mu\nu}^a G^{a \, \mu\nu}$ term. In the following we use the same notation 
as \cite{61,62}. In the limit as $\Nf$~$\to$~$\infty$ the critical propagators 
behave as 
\begin{eqnarray}
\langle \psi(p) \bar{\psi}(-p) \rangle &\sim& 
\frac{A\pslash}{(p^2)^{\mu-\tilde{\alpha}}} ~~,~~
\langle A^a_\mu(p) A^b_\nu(-p) \rangle ~\sim~ 
\frac{B\delta^{ab}}{(p^2)^{\mu-\beta}}
\left[ \eta_{\mu\nu} - \frac{p_\mu p_\nu}{p^2} \right] \nonumber \\
\langle c^a(p) \bar{c}^b(-p) \rangle &\sim& 
\frac{C\delta^{ab}}{(p^2)^{\mu-\gamma}} 
\label{largenprop}
\end{eqnarray}
in the Landau gauge. These are the dominant scaling forms of the respective
propagators. It is possible to include corrections to scaling but we omit them
here, \cite{62}. The powers of the momenta in each propagator in 
(\ref{largenprop}) are the scaling dimensions of the fields and are defined as 
\begin{equation}
\tilde{\alpha} ~=~ \mu ~-~ 1 ~+~ \half \eta ~~,~~ 
\beta ~=~ 1 ~-~ \eta ~-~ \chi ~~,~~ \gamma ~=~ \mu ~-~ 1 ~+~ \half \eta_c
\end{equation}
where $d$~$=$~$2\mu$ and $\eta$, $\chi$ and $\eta_c$ are the critical exponents
associated with the quark wave function, quark-gluon vertex operator and the 
Faddeev-Popov ghost wave function renormalization group functions. On notation 
we use $\tilde{\alpha}$ here to avoid confusion with the gauge parameter 
$\alpha$ which was the notation for the quark dimension in the early large
$\Nf$ work, \cite{61}. The remaining parts of the scaling dimensions are the 
canonical dimensions of the fields as dictated by requiring that the action is 
dimensionless in $d$-dimensions. Appending the ghost sector as discussed 
earlier there is also a ghost-gluon vertex operator anomalous dimension 
exponent $\chi_c$. However, it is not independent due to the Slavnov-Taylor 
identity. Its manifestation at the critical point requires that, \cite{61}, 
\begin{equation}
\eta_c ~=~ \eta ~+~ \chi ~-~ \chi_c
\end{equation}
is satisfied. Although at leading order in $1/\Nf$ there are no quark 
contributions to the ghost-gluon vertex and thus $\chi_{c\,1}$~$=$~$0$. We use 
the notation that an exponent, such as $\eta$, is expanded as
\begin{equation}
\eta ~=~ \sum_{i=1}^\infty \frac{\eta_i}{T_F^i\Nf^i} ~.
\end{equation}
One point concerning $\Nf$ worth noting here rests in the conventions for the 
trace over $\gamma$-matrices. Throughout we take $\mbox{Tr} I$~$=$~$4$ and 
retain four dimensional $\gamma$-matrices in six as well as two dimensions. 
This is partly because our comparison in large $\Nf$ is primarily with four 
dimensional results which use this convention and the fact that we have 
retained that convention in our six dimensional perturbative computations. One 
could of course have used higher dimensional representations for six 
dimensional $\gamma$-matrices. However, that convention can be accommodated by 
scaling $\Nf$ itself by the appropriate factor since a closed quark loop is 
always associated with a $\gamma$-matrix trace. The quantities $A$, $B$ and $C$
are the associated momentum independent amplitudes of the theory. While they 
can be evaluated in the large $\Nf$ expansion they are not central to the 
present review. 

As we will be using our results to check with the known large $\Nf$ exponents, 
it is worth collecting their values for completeness here. First, the quark 
wave function exponent $\eta$ is given by, \cite{61},
\begin{equation}
\eta_1 ~=~ C_F \eta^{\mbox{o}}_1 
\end{equation}
and, \cite{63},
\begin{eqnarray}
\eta_2 &=& \left[ \frac{2(\mu-1)(\mu-3)}{\mu(\mu-2)} ~+~ 
3\mu \left[ \Theta(\mu) - \frac{1}{(\mu-1)^2} \right] \right]
\frac{(\mu-1) C_F^2 {\eta^{\mbox{o}}_1}^2}{(\mu-2)(2\mu-1)} \nonumber \\
&& ~+ \left[
\frac{(12\mu^4-72\mu^3+126\mu^2-75\mu+11)}{2(2\mu-1)^2(2\mu-3)(\mu-2)^2} ~-~
\frac{\mu(\mu-1)}{2(2\mu-1)(\mu-2)} \left[ \Psi(\mu)^2 + \Phi(\mu) \right]
\right. \nonumber \\
&& \left. ~~~~~+ \frac{(8\mu^5-92\mu^4+270\mu^3-301\mu^2+124\mu-12) \Psi(\mu)}
{4(2\mu-1)^2(2\mu-3)(\mu-2)^2} \right] C_F C_A {\eta^{\mbox{o}}_1}^2
\end{eqnarray}
where
\begin{equation}
\eta^{\mbox{o}}_1 ~=~ -~ \frac{(2\mu-1)(2-\mu) \Gamma(2\mu)}
{4 \mu \Gamma(2-\mu) \Gamma^3(\mu)} ~.
\end{equation}
We have only provided the Landau gauge expressions since that is a fixed point
of the renormalization group functions and the large $\Nf$ arbitrary gauge
dependent expression has no relation to the critical point renormalization 
group functions for $\alpha$~$\neq$~$0$. We have defined
\begin{eqnarray}
\Theta(\mu) &=& \psi^\prime(\mu-1) ~-~ \psi^\prime(1) \nonumber \\
\Psi(\mu) &=& \psi(2\mu-3) ~+~ \psi(3-\mu) ~-~ \psi(1) ~-~ \psi(\mu-1) 
\nonumber\\
\Phi(\mu) &=& \psi^\prime(2\mu-3) ~-~ \psi^\prime(3-\mu) ~-~
\psi^\prime(\mu-1) ~+~ \psi^\prime(1) 
\end{eqnarray}
where $\psi(z)$~$=$~$\frac{d~}{dz} \ln \Gamma(z)$. At leading order the gluon
and ghost critical exponents are equivalent and are, \cite{61},
\begin{equation}
\eta ~+~ \chi ~=~ \eta_c ~=~ -~ \frac{C_A\eta^{\mbox{o}}_1}{2(\mu-2)T_F\Nf} ~+~
O \left( \frac{1}{T_F^2\Nf^2} \right) 
\end{equation}
The remaining main exponents of interest here are both gauge parameter 
independent but were evaluated in critical point large $\Nf$ using a scaling
propagator with a non-zero gauge parameter. The first such exponent is the
correction to scaling exponent $\omega$ which is the anomalous dimension of the
operator $G_{\mu\nu}^a G^{a \, \mu\nu}$. In other words $\omega$ relates to the
$\beta$-function of QCD and is the critical slope at the Wilson-Fisher fixed
point. We have \cite{62}
\begin{eqnarray}
\omega &=& (\mu - 2) ~-~ \left[ (2\mu-3)(\mu-3) C_F
- \frac{(4\mu^4 - 18\mu^3 + 44\mu^2 - 45\mu + 14) C_A }{4(2\mu-1)(\mu-1)}
\right] \frac{\eta^{\mbox{o}}_1}{T_F \Nf} \nonumber \\
&& +~ O \left( \frac{1}{T_F^2\Nf^2} \right) 
\label{qcdomega}
\end{eqnarray}
where the Quantum Electrodynamics (QED) piece was determined in \cite{102}. 
Finally, the quark mass anomalous dimension is available to two orders in large
$\Nf$ and is, \cite{63}, 
\begin{equation}
\eta_{\bar{\psi}\psi \, 1} ~=~ -~ \frac{2 C_F \eta^{\mbox{o}}_1}{(\mu-2)} 
\end{equation}
and
\begin{equation}
\eta_{\bar{\psi}\psi \, 2} ~=~ -~ \frac{2\eta_2}{(\mu-2)} ~-~ 
\frac{2(2\mu^2-4\mu+1) C_F^2 {\eta^{\mbox{o}}_1}^2}{(\mu-2)^3(2\mu-1)} ~+~ 
\frac{\mu^2(2\mu-3)^2 C_F C_A 
{\eta^{\mbox{o}}_1}^2}{4(\mu-2)^3(\mu-1)(2\mu-1)} 
\end{equation}
where $\eta_2$ was given earlier. We note that when these exponents are
expanded in $d$~$=$~$4$~$-$~$2\epsilon$ dimensions they are in agreement with
all currently available QCD renormalization group functions. This in essence is
four loops but also includes the recent five loop $\MSbar$ quark mass anomalous
dimension of \cite{103}. While $\omega$ corresponds to the gluonic operator of 
$L^{(4)}$ the exponent for the gluonic operator of $L^{(2)}$ is not independent
in the Landau gauge. This is because of a Slavnov-Taylor identity \cite{104} 
which means that the anomalous dimension of
${\cal O}$~$=$~$\half A^a_\mu A^{a\,\mu}$ is the sum of the gluon and ghost 
anomalous dimensions. This has been verified in the Landau gauge in the large 
$\Nf$ expansion, \cite{105}, and in the exponent language at leading order in 
large $\Nf$ corresponds to 
\begin{equation}
\eta_{{\cal O} \, 1} ~=~ \eta_1 ~+~ \chi_1 ~-~ \frac{1}{2} \eta_{c \, 1} ~.
\end{equation} 
While our focus here will mainly be on even dimensions the large $\Nf$ 
exponents provide information on the odd dimension versions of non-abelian
gauge theories. For instance, in five dimensions the above exponents evaluate
to 
\begin{eqnarray}
\eta &=& -~ \frac{256 C_F}{45\pi^2 T_F \Nf} ~+~ \left[ 200 \pi^2 C_A
+ 600 \pi^2 C_F - 335 C_A - 8288 C_F \right]
\frac{1024 C_F}{10125\pi^4 T_F^2 \Nf^2} \nonumber \\
&& +~ O \left( \frac{1}{T_F^3 \Nf^3} \right) \nonumber \\
\eta ~+~ \chi &=& \eta_c ~=~ \frac{256 C_A}{45\pi^2 T_F \Nf} ~+~ 
O \left( \frac{1}{T_F^2 \Nf^2} \right) \nonumber \\
\omega &=& \frac{1}{2} ~-~ \left[ 48 C_F + 103 C_A \right] 
\frac{16}{135 \pi^2 T_F \Nf} ~+~ 
O \left( \frac{1}{T_F^2 \Nf^2} \right) \nonumber \\
\eta_{\bar{\psi}\psi} &=& \frac{1024 C_F}{45\pi^2 T_F \Nf} ~-~ 
\left[ 600 \pi^2 C_A + 1800 \pi^2 C_F - 3005 C_A - 21504 C_F \right]
\frac{4096 C_F}{30375\pi^4 T_F^2 \Nf^2} \nonumber \\
&& +~ O \left( \frac{1}{T_F^3 \Nf^3} \right) ~.
\end{eqnarray}
These expressions will be of interest to any future conformal bootstrap
analysis of higher dimensional gauge theories.

\sect{Six dimensional QCD.}

We now turn to the renormalization of (\ref{lagqcd6}) at two loops in the 
$\MSbar$ scheme. This required the renormalization of the three fields and two
coupling constants. For the respective $2$- and $3$-point functions the graphs
were generated by {\sc Qgraf} and the numbers of Feynman diagrams for each are 
given in Table $1$. Compared to the corresponding renormalization in four
dimensions the numbers of graphs is similar. The main difference is in the 
triple gluon vertex renormalization due to the presence of the quintic gluon 
vertex which first arises at two loops. The sextic gluon vertex will not be 
present until three loops. Unlike the parallel eight dimensional scalar theory 
which mimics (\ref{lagqcd6}) in some ways, we do not have to consider $4$-point
vertex functions to complete the full renormalization. For each of the $2$- and
$3$-point functions we follow the same methodology and apply the Laporta 
algorithm as implemented in {\sc Reduze}. The main difference with the scalar 
theory is the presence of numerator scalar products and tensor integrals. For 
the latter we follow the projection method for the three $3$-point vertex
renormalization outlined in \cite{106}. In other words we compute the $3$-point
functions at a symmetric point where there is no nullification of external
legs. It is important to be clear why we took this more involved route. In the 
renormalization of four dimensional QCD the coupling constant renormalization 
can be deduced from a $3$-point vertex by setting an external momentum to zero.
This is an infrared safe procedure for this exceptional momentum configuration 
due to the presence of momenta in the numerator of the integrand. Moreover, 
this reduction of the computation to effectively a $2$-point function analysis 
means that the evaluation of the Feynman graphs can be computed relatively 
quickly. For (\ref{lagqcd6}) this nullification technique cannot be applied 
because the gluon and ghost propagators have higher order poles. Therefore, at 
a nullification a Feynman integral will potentially have a $1/(k^2)^4$ factor 
like (\ref{lag8}) but this cannot be infrared protected by numerator momenta in
the six dimensional gauge theory unlike the four dimensional case. In other 
words such a nullification for (\ref{lagqcd6}) would require an infrared 
rearrangement. Therefore, we have proceeded by considering each $3$-point 
function at a non-exceptional momentum configuration which is infrared safe. 
Therefore we use the same decomposition and projection of each $3$-point vertex 
into the basis of Lorentz tensors given in $d$-dimensions in \cite{106}. For 
the quark and ghost vertices the process is similar to the four dimensional 
case and does not deserve further comment. The complication occurs with the 
triple gluon vertex. From (\ref{lagqcd6}) $g_2$ only appears in the Feynman 
rules for the gluon vertices. Therefore, it might be tempting to focus on the 
renormalization of $g_2$ solely from the triple gluon vertex and assume that 
the renormalization constant for $g_1$ is deduced from one of the other two 
vertices. However, as an independent check on our {\sc Form} code and for 
completeness we have checked that the {\em same} $\MSbar$ renormalization 
constant for $g_1$ emerges from {\em each} of the three $3$-point functions. 
Moreover, we have performed the computation in an arbitrary linear covariant 
gauge and verified that $\alpha$ is absent in the $\beta$-functions. This 
non-trivial check gives us confidence in the final expressions of the 
renormalization group functions. 
 
{\begin{table}[ht]
\begin{center}
\begin{tabular}{|c||c|c|c|c|}
\hline
Green's function & One loop & Two loop & Total \\
\hline
$A^a_\mu \, A^b_\nu$ & $~\,3$ & $~\,18$ & $~\,21$ \\
$c^a \bar{c}^b$ & $~\,1$ & $~~\,\,6$ & $~~\,\,7$ \\
$\psi \bar{\psi}$ & $~\,1$ & $~~\,\,6$ & $~~\,\,7$ \\
$ A^a_\mu \, A^b_\nu A^c_\sigma$ & $~\,8$ & $115$ & $123$ \\
$c^a \bar{c}^b A^c_\sigma$ & $~\,2$ & $~\,33$ & $~\,35$ \\
$\psi \bar{\psi} A^c_\sigma$ & $~\,2$ & $~\,33$ & $~\,35$ \\
\hline
Total & $17$ & $211$ & $228$ \\
\hline
\end{tabular}
\end{center}
\begin{center}
{Table 1. Number of Feynman diagrams computed for each $2$- and $3$-point 
function.}
\end{center}
\end{table}}

The outcome of the computation is the renormalization group functions 
\begin{eqnarray}
\gamma_A(g_1,g_2,\alpha) &=& \left[ 20 \alpha C_A - 199 C_A 
- 16 \Nf T_F \right] \frac{g_1^2}{60} \nonumber \\
&& +~ \left[ 130 \alpha^2 C_A^2 g_1^3 + 1095 \alpha C_A^2 g_1^3 
- 81412 C_A^2 g_1^3 + 2178 C_A^2 g_1^2 g_2 + 5658 C_A^2 g_1 g_2^2
\right. \nonumber \\
&& \left. ~~~- 630 C_A^2 g_2^3 - 1568 C_A \Nf T_F g_1^3 
- 1248 C_A \Nf T_F g_1^2 g_2 + 192 C_A \Nf T_F g_1 g_2^2 \right. \nonumber \\
&& \left. ~~~- 6080 C_F \Nf T_F g_1^3 \right] \frac{g_1}{4320} ~+~ 
O(g_i^6) \nonumber \\
\gamma_c(g_1,g_2,\alpha) &=& C_A \left[ \alpha - 5 \right] \frac{g_1^2}{12} 
\nonumber \\ 
&& +~ \left[ - 55 \alpha^2 C_A g_1^2 + 60 \alpha C_A g_1^2 - 19952 C_A g_1^2 
+ 2700 C_A g_1 g_2 + 600 C_A g_2^2 \right. \nonumber \\
&& \left. ~~~- 1088 \Nf T_F g_1^2 \right] \frac{C_A g_1^2}{8640} ~+~ 
O(g_i^6) \nonumber \\
\gamma_\psi(g_1,g_2,\alpha) &=& C_F \left[ 3 \alpha + 5 \right] \frac{g_1^2}{6} 
\nonumber \\
&& +~ \left[ 75 \alpha^2 C_A g_1^2 + 1830 \alpha C_A g_1^2 + 43617 C_A g_1^2 
- 600 C_A g_2^2 - 8000 C_F g_1^2 \right. \nonumber \\
&& \left. ~~~+ 2048 \Nf T_F g_1^2 \right] \frac{C_F g_1^2}{4320} ~+~ O(g_i^6) 
\label{rgeqcd6}
\end{eqnarray}
for the wave function renormalization. In our convention the 
non-renormalization of $\alpha$ manifests itself in the relation 
\begin{equation}
\gamma_A(g_1,g_2,\alpha) ~+~ \gamma_\alpha(g_1,g_2,\alpha) ~=~ 0
\end{equation}
which we have checked is satisfied at two loops. The $\beta$-functions are 
\begin{eqnarray}
\beta_1(g_1,g_2) &=& \left[ - 249 C_A - 16 \Nf T_F \right] \frac{g_1^3}{120} 
\nonumber \\
&& +~ \left[ - 50682 C_A^2 g_1^3 + 2439 C_A^2 g_1^2 g_2 + 3129 C_A^2 g_1 g_2^2 
- 315 C_A^2 g_2^3 - 1328 C_A \Nf T_F g_1^3 \right. \nonumber \\
&& \left. ~~~~- 624 C_A \Nf T_F g_1^2 g_2 + 96 C_A \Nf T_F g_1 g_2^2 
- 3040 C_F \Nf T_F g_1^3 \right] \frac{g_1^2}{4320} ~+~ O(g_i^7)
\nonumber \\
\beta_2(g_1,g_2) &=& \left[ 81 C_A g_1^3 - 552 C_A g_1^2 g_2 
+ 135 C_A g_1 g_2^2 - 15 C_A g_2^3 + 104 \Nf T_F g_1^3 
- 48 \Nf T_F g_1^2 g_2 \right] \frac{1}{120} \nonumber \\
&& +~ \left[ 10212 C_A^2 g_1^5 
- 417024 C_A^2 g_1^4 g_2 
+ 142617 C_A^2 g_1^3 g_2^2
- 1014 C_A^2 g_1^2 g_2^3 
- 4725 C_A^2 g_1 g_2^4 \right. \nonumber \\
&& \left. ~~~~ 
+~ 450 C_A^2 g_2^5
- 7052 \Nf T_F C_A g_1^5 
- 20296 \Nf T_F C_A g_1^4 g_2 
+ 8868 C_A \Nf T_F g_1^3 g_2^2 \right. \nonumber \\
&& \left. ~~~~
-~ 1056 C_A \Nf T_F g_1^2 g_2^3 
+ 61600 \Nf T_F C_F g_1^5 
- 30400 \Nf T_F C_F g_1^4 g_2 \right]
\frac{1}{14400} \nonumber \\
&& +~ O(g_i^7) ~.
\label{betaqcd6}
\end{eqnarray}
The one loop term of $\beta_1(g_1,g_2)$ is clearly negative for all $\Nf$
unlike four dimensions and therefore in six dimensions the quark-gluon coupling
is asymptotically free. The corresponding one loop result in six dimensional 
QED was recently given in \cite{65} with which we agree. 

In order to provide more checks on the connection of (\ref{lagqcd6}) with lower
dimensional gauge theories at the Wilson-Fisher fixed point we have also
computed the quark mass operator anomalous dimension at two loops. To do this
we inserted the mass operator $\bar{\psi}\psi$ in a quark $2$-point function 
but such that there is a momentum flowing into the operator itself similar to 
the parallel scalar theory calculation. At one and two loops there are $1$ and 
$13$ graphs respectively. For $\alpha$~$\neq$~$0$ we find a gauge parameter 
independent $\MSbar$ expression for the quark mass operator anomalous dimension
since 
\begin{eqnarray}
\gamma_{\bar{\psi}\psi}(g_1,g_2) &=& -~ \frac{5}{3} C_F g_1^2 ~+~ 
\left[ - 11301 C_A g_1^2 + 300 C_A g_2^2 - 200 C_F g_1^2 - 544 \Nf T_F g_1^2 
\right] \frac{C_F g_1^2}{1080} \nonumber \\
&& +~ O(g_i^6) 
\end{eqnarray}
This expression was derived from the massless version of six dimensional QCD,
(\ref{lagqcd6}). As it is possible to include lower dimensional operators with
associated masses, we have also determined the renormalization of $m_i$ in 
(\ref{lagqcd6m}) at one loop in the Landau gauge. This choice of gauge is
motivated by the potential connection with the infrared structure in four
dimensions. In this instance the presence of four mass terms means that we have
to determine the mixing matrix of mass anomalous dimensions which limits this 
analysis to the leading order. However, that is sufficient to form a picture of
the how the masses relate under renormalization. If we formally label the 
operators by the label of the associated mass as given in (\ref{lagqcd6m}) then
we find that the mixing matrix, $\gamma_{ij}(g_1,g_2,\alpha)$, is sparse at one
loop and the only non-zero elements are
\begin{eqnarray}
\gamma_{11}(g_1,g_2,0) &=& -~ \frac{5}{3} C_F g_1^2 ~+~ O(g_i^4) 
\nonumber \\
\gamma_{21}(g_1,g_2,0) &=& \frac{4}{3} T_F \Nf g_1^2 ~+~ O(g_i^4) 
\nonumber \\
\gamma_{22}(g_1,g_2,0) &=& -~ \frac{2}{3} C_A g_2^2 ~-~ 2 C_A g_1 g_2 ~-~
 \frac{4}{15} T_F \Nf g_1^2 ~+~ \frac{281}{60} C_A g_1^2 ~+~ O(g_i^4) 
\nonumber \\
\gamma_{44}(g_1,g_2,0) &=& -~ \frac{2}{15} T_F \Nf g_1^2 ~-~ 
\frac{28}{15} C_A g_1^2 ~+~ O(g_i^4) ~.
\end{eqnarray}
One feature of the result is that $\gamma_{44}(g_1,g_2,0)$ satisfies 
\begin{equation}
\gamma_{44}(g_1,g_2,0) ~=~ \frac{1}{2} \left[ \gamma_A(g_1,g_2,0) ~+~ 
\gamma_c(g_1,g_2,0) \right] ~+~ O(g_i^4) ~.
\label{qcd6mixmat}
\end{equation}
parallel to the corresponding four dimensional relation. This six dimensional
result is consistent with the large $\Nf$ exponent.

One of the motivations of studying (\ref{lagqcd6}) is to establish the 
connection of four dimensional QCD with a higher dimensional gauge theory in 
the Wilson-Fisher chain. To access the large $\Nf$ exponents of the previous
section we set $d$~$=$~$6$~$-$~$2\epsilon$ and follow the algorithm of
\cite{17}. First, we define scaled couplings by
\begin{equation}
g_1 ~=~ \frac{i}{2} \sqrt{\frac{15\epsilon}{T_F\Nf}} x ~~~,~~~
g_2 ~=~ \frac{i}{2} \sqrt{\frac{15\epsilon}{T_F\Nf}} y
\end{equation}
and solve $\beta_i(g_1,g_2)$~$=$~$0$ for the critical values of $x$ and $y$ to 
$O(\epsilon^2)$. We find  
\begin{eqnarray}
x &=& 1 ~+~ \left[ -~ \frac{249}{32} C_A + \left[ \frac{475}{48} C_F
+ \frac{5855}{768} C_A \right] \epsilon \right] \frac{1}{T_F\Nf} \nonumber \\
&& +~ \left[ \frac{186003}{2048} C_A^2 + \left[ - \frac{197125}{512} C_A C_F 
- \frac{7530655}{32768} C_A^2 \right] \epsilon \right] 
\frac{1}{T_F^2\Nf^2} ~+~ O \left( \epsilon^2; \frac{1}{T_F^3\Nf^3} \right) 
\nonumber \\
y &=& \frac{13}{4} ~+~ \left[ -~ \frac{51327}{2048} C_A 
+ \left[ \frac{2325}{64} C_F + \frac{62385}{4096} C_A \right] \epsilon \right] 
\frac{1}{T_F\Nf} ~+~ O \left( \epsilon^2; \frac{1}{T_F^2\Nf^2} \right) ~. 
\end{eqnarray}
Equipped with these we have expanded out the other renormalization group 
functions, (\ref{rgeqcd6}), to the same orders as the available exponents in 
the Landau in both $\epsilon$ and $1/\Nf$ and found full agreement. Another 
check derives from $\gamma_{22}(g_1,g_2,0)$ of (\ref{qcd6mixmat}) which 
corresponds to the renormalization of the mass associated with the field 
strength operator in (\ref{lagqcd6}). In four dimensions this operator would be
the gluon kinetic term and its large $\Nf$ critical exponent, $\omega$, relates
to the critical slope of the four dimensional QCD $\beta$-function. Expanding
$\omega_1$ in (\ref{qcdomega}) to $O(\epsilon)$ near six dimensions we get 
precise agreement. For this element of (\ref{qcd6mixmat}) at one loop it will 
be an eigen-anomalous dimension as there are no other entries at this order in 
the matrix. At higher order mixing with this operator would require 
diagonalizing (\ref{qcd6mixmat}). The agreement of the perturbative results 
with the large $\Nf$ exponents is important for various reasons. For instance, 
it demonstrates that the role of the spectator operator with coupling $g_2$ is 
crucial in getting agreement. For instance, the presence of $g_2$ in 
$\gamma_{22}(g_1,g_2,0)$ is necessary for the check with $\omega$ to work at 
leading order in $\epsilon$. This spectator operator is present to ensure 
renormalizability in six dimensions but would be irrelevant in lower dimensions
at the Gaussian fixed point. That the exponents derived in the large $\Nf$ 
expansion from a critical theory with only a quark-gluon interaction is 
remarkable in some sense. Moreover it substantiates the point of view of 
\cite{60} that the triple and higher leg gluon interactions derive from 
$3$-point and higher Green's functions with only quark loops and no gluon 
interactions. That this picture extends to six dimensions establishes the same 
point of view for quintic gluon interactions in (\ref{lagqcd6}). 

Having established the connection with a lower dimensional gauge theory we now
turn to the analysis of the six dimensional renormalization group functions in
their own right. One of the interests in higher dimensional cubic scalar 
theories was to ascertain where the conformal window existed if present at all.
In four dimensional QCD this equates to the range of $\Nf$ for which a 
Banks-Zaks fixed point is present, \cite{1}. Therefore we proceed by solving
\begin{equation}
\beta_1(g_1,g_2) ~=~ \beta_2(g_1,g_2) ~=~ 0 ~~,~~ 
\frac{\partial \beta_1}{\partial g_1} \frac{\partial \beta_2}{\partial g_2}
~-~ \frac{\partial \beta_1}{\partial g_2}
\frac{\partial \beta_2}{\partial g_1} ~=~ 0 ~.
\end{equation} 
The first two determine the location of zeroes of the $\beta$-functions while 
the third is the condition for a zero eigenvalue in the matrix of 
$\beta$-function slopes. Like \cite{17} we find three solutions one of which is
real solutions and two other two are complex conjugates. The real solution is
\begin{eqnarray}
N_{\!f\,(A)} &=& 2.797566 \frac{C_A}{T_F} ~+~ 
\left[ 2.198165 C_F - 3.432003 C_A \right] \frac{\epsilon}{T_F} ~+~ 
O(\epsilon^2) \nonumber \\
x_{(A)} &=& 0.390349 ~+~ 
\left[ 0.162047 C_F - 0.064751 C_A \right] \frac{\epsilon}{C_A} ~+~ 
O(\epsilon^2) \nonumber \\
y_{(A)} &=& 0.965498 ~+~ 
\left[ 0.412927 C_A + 0.185332 C_F \right] \frac{\epsilon}{C_A} ~+~ 
O(\epsilon^2) ~.
\end{eqnarray}
and the other two are 
\begin{eqnarray}
N_{\!(B)} &=& 
\left[ 3.283595 + 0.660678 i \right] \frac{C_A}{T_F} \nonumber \\
&& +~ \left[ [ - 2.089275 - 3.907327 i ] C_A
+ [ 3.737235 + 1.869896 i] C_F \right] \frac{\epsilon}{T_F} ~+~ O(\epsilon^2)
\nonumber \\
x_{(B)} &=& 0.0344173 i + 0.420036 \nonumber \\
&& +~ \left[ [ - 0.039289 - 0.150826 i ] C_A 
+ [ 0.244905 + 0.074815 i ] C_F \right] \frac{\epsilon}{C_A} ~+~ O(\epsilon^2)
\nonumber \\
y_{(B)} &=& 1.467391 - 0.100116 i \nonumber \\
&& +~ \left[ [ - 0.079478 - 0.487643 i ] C_A 
+ [ 1.077389 - 0.083226 i ] C_F \right] \frac{\epsilon}{C_A} ~+~ O(\epsilon^2)
\end{eqnarray}
and its complex conjugate denoted by $C$. For reference the real solution for 
$SU(3)$ is  
\begin{eqnarray}
\left. N_{\!f\,(A)} \right|_{SU(3)} &=& 16.785398 ~-~ 14.730246 \epsilon ~+~ 
O(\epsilon^2) \nonumber \\
\left. x_{(A)} \right|_{SU(3)} &=& 0.390349 ~+~ 0.007270 \epsilon ~+~ 
O(\epsilon^2) \nonumber \\
\left. y_{(A)} \right|_{SU(3)} &=& 0.965498 ~+~ 0.495297 \epsilon ~+~ 
O(\epsilon^2) ~. 
\label{critnfsu3}
\end{eqnarray}
Interestingly the location of the conformal window in purely six dimensions is 
between $\Nf$~$=$~$16$ and $17$ similar to four dimensional QCD. However, in 
\cite{17} the higher dimensional theory and the $\epsilon$ expansion was used 
to estimate the boundary of the window in a lower dimension by summation. If we 
consider that approach the two loop correction to the real solution, 
$N_{\!f\,(A)}$, is comparable to the one loop part. This suggests that 
perturbation theory may not be reliable. However, using a simple Pad\'{e} 
approximant, which is possible due to the negative correction, then in four 
dimensions we find $N_{\!f\,(A)}$~$=$~$8.939991$. This is lower than the 
leading order, and similar to the situation in scalar $O(N)$ $\phi^3$ theory. 
It would be interesting to see what effect the three loop corrections would 
have on this critical $\Nf$ value. 
 
Having found the region where there is a conformal window it is worth analysing
the renormalization group functions within this for specified values of $\Nf$. 
We take $\Nf$~$=$~$3$, $12$ and $16$. These choices are motivated by values in 
four dimensions. For instance, the first is because it corresponds to the 
number of light quarks. The value of $16$ is chosen since it is the largest
within the six dimensional conformal window. Finally, we consider 
$\Nf$~$=$~$12$ since there is interest in four dimensional theories with this
value due to trying to understand the Banks-Zaks fixed point non-perturbatively
on the lattice. For each of the cases there are four solutions to the equations
\begin{equation}
\beta_1(g_1,g_2) ~=~ 0 ~~,~~ \beta_2(g_1,g_2) ~=~ 0 
\end{equation}
for a particular value of $\Nf$ excluding the trivial one. In this counting we
ignore solutions which are obtained from these by reflections 
$g_i$~$\to$~$-$~$g_i$. For each value of $\Nf$ we give the location of the 
fixed point in terms of $x$ and $y$ and the renormalization group functions 
evaluated at each fixed point in the Landau gauge. Included in this are the 
eigen-critical exponents $\omega_\pm$ which are the eigenvalues of the matrix 
$\frac{\partial \beta_i}{\partial g_j}$. The signs of these exponents determine
the stability or otherwise of the fixed point. In our labelling of the four 
non-trivial solutions for each $\Nf$ value, solutions $1$ and $3$ are stable 
while $2$ and $4$ are saddle points. For solutions $1$, $2$ and $3$ the first 
term of the same exponent in each solution is the same. This is because the one
loop term of the corresponding renormalization group function only depends on 
$g_1$ and there is no $g_2$ dependence in the one loop term of 
$\beta_1(g_1,g_2)$. The values for the exponents begin to differ at 
$O(\epsilon^2)$ due to $g_2$ appearing in the two loop expressions. This is the
main reason for a two loop renormalization as a one loop analysis would not
reveal distinctive differences. The solution labelled $4$ is somewhat different
in that it corresponds to $g_1$~$=$~$0$. In effect there are no quarks or 
Faddeev-Popov ghosts in the corresponding Lagrangian but only the $3$-leg gauge
invariant operator. Also the kinetic term for the gluon derives from the free 
part of ${\cal O}^{(6)}_1$. In some sense this solution is not interesting as 
all the critical exponents are zero except $\omega_\pm$ which take their 
canonical values of $2\epsilon$ and $-$~$\epsilon$ respectively. Therefore, 
solution $4$ appears to be in effect a free field solution. So we do not 
explicitly record any exponent values for solution $4$.

More specifically our results for $\Nf$~$=$~$3$ are
\begin{eqnarray}
x_{(1)} &=& 0.176432 ~-~ 0.003730 \epsilon ~+~ O(\epsilon^2) \nonumber \\
y_{(1)} &=& 0.936586 ~+~ 0.548703 \epsilon ~+~ O(\epsilon^2) \nonumber \\
\left. \gamma_A(g_1,g_2,0) \right|_{(1)} &=&
0.805447 \epsilon ~-~ 0.127340 \epsilon^2 ~+ O(\epsilon^2) \nonumber \\
\left. \gamma_c(g_1,g_2,0) \right|_{(1)} &=&
0.097276 \epsilon ~+~ 0.063670 \epsilon^2 ~+~ O(\epsilon^2) \nonumber \\
\left. \gamma_\psi(g_1,g_2,0) \right|_{(1)} &=&
-~ 0.086468 \epsilon ~+~ 0.139232 \epsilon^2 ~+~ O(\epsilon^2) \nonumber \\
\left. \gamma_{\bar{\psi}\psi}(g_1,g_2,0) \right|_{(1)} &=&
0.172936 \epsilon ~-~ 0.079266 \epsilon^2 ~+~ O(\epsilon^2) \nonumber \\
\left. \omega_+ \right|_{(1)} &=& 2.000000 \epsilon ~+~ 
0.084558 \epsilon^2 ~+~ O(\epsilon^3) \nonumber \\
\left. \omega_- \right|_{(1)} &=& 0.598467 \epsilon ~+~ 
0.971180 \epsilon^2 ~+~ O(\epsilon^3) \nonumber \\
x_{(2)} &=& 0.176432 ~+~ 0.052245 \epsilon ~+~ O(\epsilon^2) \nonumber \\
y_{(2)} &=& 0.558153 ~+~ 0.113962 \epsilon ~+~ O(\epsilon^2) \nonumber \\
\left. \gamma_A(g_1,g_2,0) \right|_{(2)} &=&
0.805447 \epsilon ~-~ 0.040157 \epsilon^2 ~+~ O(\epsilon^2) \nonumber \\
\left. \gamma_c(g_1,g_2,0) \right|_{(2)} &=&
0.097276 \epsilon ~+~ 0.020079 \epsilon^2 ~+~ O(\epsilon^2) \nonumber \\
\left. \gamma_\psi(g_1,g_2,0) \right|_{(2)} &=&
-~ 0.086468 \epsilon ~+~ 0.145505 \epsilon^2 ~+~ O(\epsilon^2) \nonumber \\
\left. \gamma_{\bar{\psi}\psi}(g_1,g_2,0) \right|_{(2)} &=&
0.172936 \epsilon ~-~ 0.091812 \epsilon^2 ~+~ O(\epsilon^2) \nonumber \\
\left. \omega_+ \right|_{(2)} &=& 2.000000 \epsilon ~-~ 
1.184488 \epsilon^2 ~+~ O(\epsilon^3) \nonumber \\
\left. \omega_- \right|_{(2)} &=& -~ 0.329946 \epsilon ~-~ 
0.055233 \epsilon^2 ~+~ O(\epsilon^3) \nonumber \\
x_{(3)} &=& 0.176432 ~+~ 0.111433 \epsilon ~+~ O(\epsilon^2) \nonumber \\
y_{(3)} &=& 0.093152 ~+~ 0.204520 \epsilon ~+~ O(\epsilon^2) \nonumber \\
\left. \gamma_A(g_1,g_2,0) \right|_{(3)} &=&
0.805447 \epsilon ~-~ 0.007256 \epsilon^2 ~+~ O(\epsilon^2) \nonumber \\
\left. \gamma_c(g_1,g_2,0) \right|_{(3)} &=&
0.097276 \epsilon ~+~ 0.003628 \epsilon^2 ~+~ O(\epsilon^2) \nonumber \\
\left. \gamma_\psi(g_1,g_2,0) \right|_{(3)} &=&
-~ 0.086468 \epsilon ~+~ 0.120224 \epsilon^2 ~+~ O(\epsilon^2) \nonumber \\
\left. \gamma_{\bar{\psi}\psi}(g_1,g_2,0) \right|_{(3)} &=&
0.172936 \epsilon ~-~ 0.041251 \epsilon^2 ~+~ O(\epsilon^2) \nonumber \\
\left. \omega_+ \right|_{(3)} &=& 2.000000 \epsilon ~-~ 
2.526371 \epsilon^2 ~+~ O(\epsilon^3) \nonumber \\
\left. \omega_- \right|_{(3)} &=& 0.735370 \epsilon ~-~ 
0.731156 \epsilon^2 ~+~ O(\epsilon^3) \nonumber \\
x_{(4)} &=& O(\epsilon^2) ~~,~~ 
y_{(4)} ~=~ 0.730297 ~-~ 0.365148 \epsilon ~+~ O(\epsilon^2) ~. 
\end{eqnarray}
When $\Nf$~$=$~$12$ we have 
\begin{eqnarray}
x_{(1)} &=& 0.337460 ~+~ 0.040482 \epsilon ~+~ O(\epsilon^2) \nonumber \\ 
y_{(1)} &=& 1.540384 ~+~ 1.051213 \epsilon ~+~ O(\epsilon^2) \nonumber \\
\left. \gamma_A(g_1,g_2,0) \right|_{(1)} &=&
0.822064 \epsilon ~-~ 0.071166 \epsilon^2 ~+~ O(\epsilon^2) \nonumber \\
\left. \gamma_c(g_1,g_2,0) \right|_{(1)} &=&
0.088968 \epsilon ~+~ 0.035583 \epsilon^2 ~+~ O(\epsilon^2) \nonumber \\
\left. \gamma_\psi(g_1,g_2,0) \right|_{(1)} &=&
-~ 0.079083 \epsilon ~+~ 0.129510 \epsilon^2 ~+~ O(\epsilon^2) \nonumber \\
\left. \gamma_{\bar{\psi}\psi}(g_1,g_2,0) \right|_{(1)} &=&
0.158165 \epsilon ~-~ 0.078886 \epsilon^2 ~+~ O(\epsilon^2) \nonumber \\
\left. \omega_+ \right|_{(1)} &=& 2.00000 \epsilon ~-~ 
0.479845 \epsilon^2 ~+~ O(\epsilon^3) \nonumber \\
\left. \omega_- \right|_{(1)} &=& 0.256816 \epsilon ~+~ 
0.215915 \epsilon^2 ~+~ O(\epsilon^3) \nonumber \\
x_{(2)} &=& 0.337460 ~+~ 0.108330 \epsilon ~+~ O(\epsilon^2) \nonumber \\ 
y_{(2)} &=& 1.030159 ~+~ 0.255942 \epsilon ~+~ O(\epsilon^2) \nonumber \\
\left. \gamma_A(g_1,g_2,0) \right|_{(2)} &=&
0.822064 \epsilon ~-~ 0.026704 \epsilon^2 ~+~ O(\epsilon^2) \nonumber \\
\left. \gamma_c(g_1,g_2,0) \right|_{(2)} &=&
0.088968 \epsilon ~+~ 0.013352 \epsilon^2 ~+~ O(\epsilon^2) \nonumber \\
\left. \gamma_\psi(g_1,g_2,0) \right|_{(2)} &=&
-~ 0.079083 \epsilon ~+~ 0.130122 \epsilon^2 ~+~ O(\epsilon^2) \nonumber \\
\left. \gamma_{\bar{\psi}\psi}(g_1,g_2,0) \right|_{(2)} &=&
0.158165 \epsilon ~-~ 0.080112 \epsilon^2 ~+~ O(\epsilon^2) \nonumber \\
\left. \omega_+ \right|_{(2)} &=& 2.000000 \epsilon ~-~ 
1.284069 \epsilon^2 ~+~ O(\epsilon^3) \nonumber \\
\left. \omega_- \right|_{(2)} &=& -~ 0.134787 \epsilon ~+~ 
0.145784 \epsilon^2 ~+~ O(\epsilon^3) \nonumber \\
x_{(3)} &=& 0.337460 ~+~ 0.174074 \epsilon ~+~ O(\epsilon^2) \nonumber \\ 
y_{(3)} &=& 0.466593 ~+~ 1.074306 \epsilon ~+~ O(\epsilon^2) \nonumber \\
\left. \gamma_A(g_1,g_2,0) \right|_{(3)} &=&
0.822064 \epsilon ~-~ 0.001544 \epsilon^2 ~+~ O(\epsilon^2) \nonumber \\
\left. \gamma_c(g_1,g_2,0) \right|_{(3)} &=&
0.088968 \epsilon ~+~ 0.000772 \epsilon^2 ~+~ O(\epsilon^2) \nonumber \\
\left. \gamma_\psi(g_1,g_2,0) \right|_{(3)} &=&
-~ 0.079083 \epsilon ~+~ 0.120155 \epsilon^2 ~+~ O(\epsilon^2) \nonumber \\
\left. \gamma_{\bar{\psi}\psi}(g_1,g_2,0) \right|_{(3)} &=&
0.158165 \epsilon ~-~ 0.060177 \epsilon^2 ~+~ O(\epsilon^2) \nonumber \\
\left. \omega_+ \right|_{(3)} &=& 2.000000 \epsilon ~-~ 
2.063346 \epsilon^2 ~+~ O(\epsilon^3) \nonumber \\
\left. \omega_- \right|_{(3)} &=& 0.283665 \epsilon ~-~ 
0.844111 \epsilon^2 ~+~ O(\epsilon^3) \nonumber \\
x_{(4)} &=& O(\epsilon^2) ~~,~~  
y_{(4)} ~=~ 1.460593 ~-~ 0.730297 \epsilon ~+~ O(\epsilon^2) ~. 
\end{eqnarray}
Finally, 
\begin{eqnarray}
x_{(1)} &=& 0.382473 ~+~ 0.070883 \epsilon ~+~ O(\epsilon^2) \nonumber \\ 
y_{(1)} &=& 1.584443 ~+~ 0.868230 \epsilon ~+~ O(\epsilon^2) \nonumber \\
\left. \gamma_A(g_1,g_2,0) \right|_{(1)} &=&
0.828571 \epsilon ~-~ 0.050107 \epsilon^2 ~+~ O(\epsilon^2) \nonumber \\
\left. \gamma_c(g_1,g_2,0) \right|_{(1)} &=&
0.085714 \epsilon ~+~ 0.025053 \epsilon^2 ~+~ O(\epsilon^2) \nonumber \\
\left. \gamma_\psi(g_1,g_2,0) \right|_{(1)} &=&
-~ 0.076190 \epsilon ~+~ 0.125124 \epsilon^2 ~+~ O(\epsilon^2) \nonumber \\
\left. \gamma_{\bar{\psi}\psi}(g_1,g_2,0) \right|_{(1)} &=&
0.152381 \epsilon ~-~ 0.077478 \epsilon^2 ~+~ O(\epsilon^2) \nonumber \\
\left. \omega_+ \right|_{(1)} &=& 2.000000 \epsilon ~-~ 
0.741312 \epsilon^2 ~+~ O(\epsilon^3) \nonumber \\
\left. \omega_- \right|_{(1)} &=& 0.144292 \epsilon ~-~ 
0.115028 \epsilon^2 ~+~ O(\epsilon^3) \nonumber \\
x_{(2)} &=& 0.382473 ~+~ 0.135377 \epsilon ~+~ O(\epsilon^2) \nonumber \\ 
y_{(2)} &=& 1.067822 ~-~ 0.219836 \epsilon ~+~ O(\epsilon^2) \nonumber \\
\left. \gamma_A(g_1,g_2,0) \right|_{(2)} &=&
0.828571 \epsilon ~-~ 0.017142 \epsilon^2 ~+~ O(\epsilon^2) \nonumber \\
\left. \gamma_c(g_1,g_2,0) \right|_{(2)} &=&
0.085714 \epsilon ~+~ 0.008571 \epsilon^2 ~+~ O(\epsilon^2) \nonumber \\
\left. \gamma_\psi(g_1,g_2,0) \right|_{(2)} &=&
-~ 0.076190 \epsilon ~+~ 0.123897 \epsilon^2 ~+~ O(\epsilon^2) \nonumber \\
\left. \gamma_{\bar{\psi}\psi}(g_1,g_2,0) \right|_{(2)} &=&
0.152381 \epsilon ~-~ 0.075023 \epsilon^2 ~+~ O(\epsilon^2) \nonumber \\
\left. \omega_+ \right|_{(2)} &=& 2.000000 \epsilon ~-~ 
1.415811 \epsilon^2 ~+~ O(\epsilon^3) \nonumber \\
\left. \omega_- \right|_{(2)} &=& -~ 0.050461 \epsilon ~+~ 
0.434552 \epsilon^2 ~+~ O(\epsilon^3) \nonumber \\
x_{(3)} &=& 0.382473 ~+~ 0.166952 \epsilon ~+~ O(\epsilon^2) \nonumber \\ 
y_{(3)} &=& 0.789993 ~+~ 2.368779 \epsilon ~+~ O(\epsilon^2) \nonumber \\
\left. \gamma_A(g_1,g_2,0) \right|_{(3)} &=&
0.828571 \epsilon ~-~ 0.005495 \epsilon^2 ~+~ O(\epsilon^2) \nonumber \\
\left. \gamma_c(g_1,g_2,0) \right|_{(3)} &=&
0.085714 \epsilon ~+~ 0.002748 \epsilon^2 ~+~ O(\epsilon^2) \nonumber \\
\left. \gamma_\psi(g_1,g_2,0) \right|_{(3)} &=&
-~ 0.076190 \epsilon ~+~ 0.120535 \epsilon^2 ~+~ O(\epsilon^2) \nonumber \\
\left. \gamma_{\bar{\psi}\psi}(g_1,g_2,0) \right|_{(3)} &=&
0.152381 \epsilon ~-~ 0.068298 \epsilon^2 ~+~ O(\epsilon^2) \nonumber \\
\left. \omega_+ \right|_{(3)} &=& 2.000000 \epsilon ~-~ 
1.746028 \epsilon^2 ~+~ O(\epsilon^3) \nonumber \\
\left. \omega_- \right|_{(3)} &=& 0.077597 \epsilon ~-~ 
0.989589 \epsilon^2 ~+~ O(\epsilon^3) \nonumber \\
x_{(4)} &=& O(\epsilon^2) ~~,~~ 
y_{(4)} ~=~ 1.686548 ~-~ 0.843274 \epsilon ~+~ O(\epsilon^2) 
\end{eqnarray}
for $\Nf$~$=$~$16$. For $\Nf$~$>$~$16$ there are two real solutions and two
complex conjugate solutions ignoring the reflection symmetry. For the real
solutions one is stable while the other is a saddle point. The former has a
non-zero value for $g_1$ at criticality and is the solution which in effect
corresponds to the large $\Nf$ solution. The other real solution is the 
effective free field solution as it corresponds to $g_1$~$=$~$0$.  

\sect{Higher dimensional QED.}

Having concentrated for the most part on non-abelian gauge theories we devote 
the remainder of our analysis to abelian theories in six and higher dimensions.
One of the reasons for this is that the analysis is more straightforward due
to fewer interactions and also because of recent activity in this area,
\cite{64,65}. The easier calculability has allowed the authors of \cite{64,65} 
to extract interesting features of the $F$-theorem in higher dimensional 
abelian gauge theories which may be shared with non-abelian ones. Based on our 
earlier considerations the six dimensional QED Lagrangian is, \cite{65}, 
\begin{equation}
\left. L^{(6)} \right|_{\mbox{\footnotesize{QED}}} ~=~ 
-~ \frac{1}{4} \left( \partial_\mu F_{\nu\sigma} \right)
\left( \partial^\mu F^{\nu\sigma} \right) ~-~
\frac{1}{2\alpha} \left( \partial_\mu \partial^\nu A_\nu \right)
\left( \partial^\mu \partial^\sigma A_\sigma \right) ~+~
i \bar{\psi}^i \Dslash \psi^i ~.
\label{lagqed6}
\end{equation}
The main differences are the absence of the $3$-point operator with coupling
$g_2$ which was proportional to the colour group structure functions and the 
replacement of the covariant derivative in the gauge field kinetic term by the 
partial derivative. The gauge fixing term is similar to QCD but in a linear 
covariant gauge there are no Faddeev-Popov ghosts. The upshot is that we have
renormalized (\ref{lagqed6}) to {\em three} loops in the $\MSbar$ scheme. We
find 
\begin{eqnarray}
\gamma_A(g_1,\alpha) &=& -~ \frac{4}{15} \Nf g_1^2 ~-~ 
\frac{38}{27} \Nf g_1^4 ~+~ 17 \Nf \left[ 200 - 111 \Nf \right] 
\frac{g_1^6}{6075} ~+~ O(g_1^8) \nonumber \\
\gamma_\psi(g_1,\alpha) &=& \left[ 3 \alpha + 5 \right] \frac{g_1^2}{6} ~+~ 
2 \left[ 32 \Nf - 125 \right] \frac{g_1^4}{135} \nonumber \\
&& +~ \left[ 2864 \Nf^2 - 648000 \zeta_3 \Nf + 730375 \Nf + 1944000 \zeta_3 
- 1033000 \right] \frac{g_1^6}{243000} \,+\, O(g_1^8) \nonumber \\
\beta_1(g_1) &=& -~ \frac{2}{15} \Nf g_1^3 ~-~ \frac{19}{27} \Nf g_1^5 ~+~
17 \Nf \left[ 200 - 111 \Nf \right] \frac{g_1^7}{12150} ~+~ O(g_1^9) 
\label{rgeqed6}
\end{eqnarray}
where we confirm the one loop asymptotically free $\beta$-function of 
\cite{65}. To derive these expressions we have independently renormalized the
photon $2$-point function and the electron-photon vertex separately so that
the Ward-Takahashi identity
\begin{equation}
\beta_1(g_1) ~=~ \frac{g_1}{2} \gamma_A(g_1,\alpha) 
\end{equation}
emerges naturally and plays the role of a computational check. One advantage of
considering the abelian theory is that there are no triple or quartic photon 
vertices. Thus we can access the vertex renormalization by the method discussed
in \cite{33}. There six dimensional $\phi^3$ theory was renormalized to four 
loops by considering only $2$-point functions. The $3$-point Green's functions 
with zero momentum insertions were generated by expanding the massless 
propagator with the appropriate Feynman rule for the insertion. Also for 
(\ref{lagqed6}) nullification does not involve a vertex with three photons. So 
no infrared problems arise which prevented us from using this approach in QCD. 
Overall this reduces the number of graphs to be evaluated. In addition we have 
also determined the $\MSbar$ electron mass anomalous dimension which is 
\begin{eqnarray}
\gamma_{\bar{\psi}\psi}(g_1) &=& -~ \frac{5 g_1^2}{3} ~+~ 
\left[ - 68 \Nf - 25 \right] \frac{g_1^4}{135} \nonumber \\
&& +~ \left[ 13456 \Nf^2 + 648000 \zeta_3 \Nf - 818575 \Nf + 1215000 \zeta_3 
- 726875 \right] \frac{g_1^6}{121500} \nonumber \\
&& +~ O(g_1^8) ~.
\end{eqnarray}
From the three loop results, there are several interesting features. First, we
have computed all renormalization group functions in terms of a non-zero
$\alpha$. However, from (\ref{rgeqed6}) the only place where $\alpha$ appears 
in these $\MSbar$ results is in the one loop term of the electron wave function
anomalous dimension. Clearly in $\MSbar$ the $\beta$-function will be $\alpha$ 
independent which by the Ward-Takahashi identity means that the photon
anomalous dimension is independent of the gauge parameter. The absence of 
$\alpha$ beyond one loop in $\gamma_\psi(g_1,\alpha)$ might be surprising if it
was not in fact completely parallel to the situation in four dimensions.
Indeed from explicit four loop computations $\alpha$ is absent after one loop,
\cite{107}. In \cite{108,109} an argument was given which suggested that to all
orders $\alpha$ appears only in the one loop term. The fact that such a 
property seems to be present in six dimensions suggests that the result is
independent of dimension. The next observation concerns the $\beta$-function 
which is that each term is negative for $\Nf$~$>$~$1$. When $\Nf$~$=$~$1$ there
is a pseudo-Banks-Zaks fixed point at $g_1$~$=$~$2.415479$. We have attributed 
it as a non-standard Banks-Zaks fixed point as it derives from an imbalance of 
the signs of the first three terms rather than the one and two loop terms in 
the QCD case. As such it may not survive in a four loop analysis. As a check on
our three loop results we have verified that the critical exponents determined 
at the Wilson-Fisher fixed point agree precisely with the corresponding ones 
determined at various orders in the large $\Nf$ expansion, \cite{61,62,102}. 
Those exponents being expanded in the neighbourhood of six dimensions. This is 
another reason why we considered the abelian theory at one loop order higher 
than the non-abelian extension. It was important to put a six dimensional gauge
theory on the same footing as scalar $\phi^3$ theory. In other words there is a
tower of gauge theories driven by a common interaction and underlying symmetry. 

One observation which has been made in the context of the tower of theories in
$d$-dimensions is that at the Wilson-Fisher fixed point one of the two
connecting theories is asymptotically free while the other is 
non-asymptotically free, \cite{65}. In other words there is a type of
ultraviolet/infrared duality across the dimensions such that the infrared fixed
point of one is an ultraviolet fixed point of the other. In the $O(N)$ scalar 
theory case the nonlinear $\sigma$ model is asymptotically free in two 
dimensions whereas $\phi^4$ theory is not in four dimensions. The six 
dimensional partner is asymptotically free. Although it is not immediately 
clear if this is the case in the eight dimensional cousin, (\ref{lag8}). This 
is because asymptotic freedom usually refers to the theory with a single scalar
field and no $O(N)$ symmetry. In the theories in six and lower dimensions they 
all have a single coupling in that instance. In the eight dimensional scalar 
theory case in the absence of the $O(N)$ symmetry there are two couplings. 
Setting $g_1$~$=$~$0$ in (\ref{rge8}) both one loop terms of 
$\beta_2(g_1,g_2,g_3)$ and $\beta_3(g_1,g_2,g_3)$ are positive. So in this 
instance it appears that the base eight dimensional theory is not 
asymptotically free. As noted in \cite{65} a similar picture is present in the 
QED tower. In four dimensions QED is not asymptotically free whereas in six 
dimensions it is, \cite{65}. As we have considered the eight dimensional $O(N)$
scalar theory extension it is worthwhile repeating the exercise for QED in 
eight dimensions. To write down the Lagrangian one has to follow our earlier 
prescription which requires extra interactions akin to the situation for 
(\ref{lag8}). We have 
\begin{eqnarray}
\left. L^{(8)} \right|_{\mbox{\footnotesize{QED}}} &=& 
-~ \frac{1}{4} \left( \partial_\mu \partial_\nu F_{\sigma\rho} \right)
\left( \partial^\mu \partial^\nu F^{\sigma\rho} \right) ~-~
\frac{1}{2\alpha} \left( \partial_\mu \partial^\nu A_\nu \right)
\left( \partial^\mu \partial^\sigma A_\sigma \right) \nonumber \\
&& +~ i \bar{\psi}^i \Dslash \psi^i ~+~ 
\frac{g_2^2}{32} F_{\mu\nu} F^{\mu\nu} F_{\sigma\rho} F^{\sigma\rho} ~+~
\frac{g_3^2}{8} F_{\mu\nu} F^{\mu\sigma} F_{\nu\rho} F^{\sigma\rho} ~.
\label{lagqed8}
\end{eqnarray}
The corresponding QCD Lagrangian would be much more involved. For instance, 
dimension eight and ten gluonic operators were considered for $SU(\Nc)$ gauge 
theories in \cite{66}. Equipped with (\ref{lagqed8}) we have found the 
renormalization group functions to a similar order as (\ref{lag8}) are
\begin{eqnarray}
\gamma_A(g_1,g_2,g_3,\alpha) &=& \frac{\Nf g_1^2}{35} ~+~ 
\frac{11 \Nf g_1^4}{120} ~+~ O(g_i^6) \nonumber \\
\gamma_\psi(g_1,g_2,g_3,\alpha) &=& \left[ 2 \alpha + 7 \right] 
\frac{g_1^2}{12} ~+~
\left[ - 964 \Nf - 13475 \right] \frac{g_1^4}{33600} ~+~ O(g_i^6) \nonumber \\
\beta_1(g_1,g_2,g_3) &=& \frac{ \Nf g_1^3}{70} ~+~ 
\frac{11 \Nf g_1^5}{240} ~+~ O(g_i^7) \nonumber \\
\beta_2(g_1,g_2,g_3) &=& \left[ 1120 g_1^4 \Nf + 72 g_1^2 g_2^2 \Nf - 861 g_2^4
- 1659 g_2^2 g_3^2 - 609 g_3^4 \right] \frac{1}{1260} ~+~ O(g_i^6) \nonumber \\
\beta_3(g_1,g_2,g_3) &=& \left[ - 1568 g_1^4 \Nf + 144 g_1^2 g_3^2 \Nf 
- 21 g_2^4 - 294 g_2^2 g_3^2 - 1029 g_3^4 \right] \frac{1}{2520} ~+~ O(g_i^6) 
\nonumber \\ 
\gamma_{\bar{\psi}\psi}(g_1,g_2,g_3) &=& 
-~ \frac{7 g_1^2}{12} ~+~ \left[ 2052 \Nf - 1225 \right] 
\frac{g_1^4}{100800} ~+~ O(g_i^6) ~.
\label{rgeqed8}
\end{eqnarray}
The structure of these functions is different from those of (\ref{lag8}). The
absence of a triple photon vertex means that at one loop there are no $g_2$ or
$g_3$ couplings in $\beta_1(g_1,g_2,g_3)$. That this persists at two loops is
somewhat surprising given that there is one topology which involves a quartic
photon vertex in the electron-photon vertex function. It transpires that the
graph is finite. Equally the two loop photonic sunset graph in the photon
$2$-point function is also finite which ensures the Ward-Takahashi identity is
not violated. The absence of $g_2$ and $g_3$ dependence at least to two loops
in $\beta_1(g_1,g_2,g_3)$ exposes the non-asymptotic freedom of eight
dimensional QED. Next we note that at least at two loops the electron anomalous
dimension has no gauge parameter dependence in the two loop term. While this is
not inconsistent with the lower dimensional observations it again lends some 
weight to the one loop $\alpha$ dependence being dimension independent. The 
final comment on (\ref{rgeqed8}) is that we have again verified that the 
critical exponents at the Wilson-Fisher fixed point are in exact agreement with
the exponents from the large $\Nf$ expansion when evaluated near eight 
dimensions. 

In light of these latter remarks it is worth making a few brief comments about
what lies beyond eight dimensions for QED. For instance, one can try and
address the issue of asymptotic freedom in higher dimensions by exploiting
properties of the gauge theory which are not present in a scalar theory. One
feature in QED is that the $\beta$-function of the gauge coupling to matter can
be deduced from the photon $2$-point function. At one loop the graph does not
involve photon propagators. This is under the assumption that there are no
triple photon vertices. If such a $3$-point vertex is present then the 
following argument will be invalid. However, if the only $3$-point vertex is
the electron-photon one then from the photon $2$-point function the one loop
$\beta$-function is  
\begin{equation}
\beta^{(D)}_1(g_1,g_2,\ldots) ~=~ 
\frac{ 2 (-1)^{D/2} \Gamma(\half D) \Nf g_1^3}{(D-1) \Gamma(D-2)} ~+~ O(g_i^5) 
\label{betaqedd}
\end{equation}
in $d$~$=$~$D$~$-$~$2\epsilon$ where $D$ is an even integer with $D$~$>$~$2$.
The expression tallies with the known results up to eight dimensions. Under
the assumption we have made it is evident that QED yoyos between being 
asymptotically free and not being asymptotically free. The origin of the
varying sign is the residue of the simple poles in the Euler $\Gamma$-function
when one expands around the appropiate simple pole in $\epsilon$ to 
determine the photon wave function renormalization constant and via the
Ward-Takahashi identity the $\beta$-function of $g_1$. In using 
(\ref{betaqedd}) it is important to realise that it is only valid for even
integers larger than two. It cannot be used in the intervening continuous
dimensions and expressed in terms of a regularizing parameter which has already
been used to determine (\ref{betaqedd}) in the $\MSbar$ scheme.

\sect{Discussion.}

We make some closing observations. First, we have achieved one of the main 
goals which was to construct and establish higher dimensional field theories 
which lie in the same universality class as already well-established theories 
at the Wilson-Fisher fixed point. The process is based on a common interaction 
which underpins each Lagrangian in a chain as well as renormalizability. Aside 
from the fields being in the same symmetry groups, one consequence is the 
appearance of extra interactions over and above the core one connecting all 
candidates. These spectator interactions play a key role in ensuring 
$d$-dimensional equivalences. In their critical dimension the extra coupling 
constants produce a rich spectrum of fixed points and if analysed in 
$d$-dimensions several of these may be connected to non-trivial and perhaps 
non-perturbative fixed points in the companion lower dimensional model. One 
hint of this, for example, may be in the infrared behaviour of the four 
dimensional gluon propagator. In the Landau gauge it has been shown to freeze 
to a finite non-zero value at zero momentum in lattice analyses over recent 
years, \cite{49,50,51,52,53,54,55,56,57,58,59}. Such behaviour for the gluon 
and Faddeev-Popov ghost propagators can be mimicked from the six dimensional 
gauge theory if one allows for the presence of lower dimensional operators in 
the Lagrangian with associated masses. While this approach should be regarded 
as a model it may be indicative that higher dimensional operators, including 
the six dimensional ones considered here, could become relevant in the critical
sense and be the dominant operators driving the gluon propagator infrared 
behaviour. There is evidence from a Schwinger-Dyson analysis to support this,
\cite{98}. On a related issue we have established that six dimensional QCD has 
asymptotic freedom. So this theory is potentially another where the issues of 
colour confinement could be investigated especially as its abelian partner is 
also asymptotically free but probably does not have confinement. At this stage 
it is still perhaps premature to think that links with lower dimensional 
non-perturbative fixed points has been fully established. This is primarily 
because in the gauge theory we only performed the renormalization to two loops.
This was mainly to demonstrate the viability of the approach. A one loop 
computation would not really have been sufficient since in the critical 
dimension the effect of the spectator interaction coupling constant does not 
appear in anomalous dimensions until two loops. Going beyond two loops is 
possible but not as straightforward as for a four dimensional gauge theory due 
to the technical issues surrounding infrared problems if vertices are computed 
at exceptional momenta configurations. However, using a non-exceptional set-up 
would require the three loop $3$-point masters which are not yet known even in 
four dimensions. With the two loop renormalization of (\ref{lagqcd6}) it should
be possible to extend the $F$-theorem studies in six dimensional QED, 
\cite{65}, to the non-abelian case. Moreover, it would be interesting to 
compare the perturbative picture with a gauge theory conformal bootstrap 
analysis. 

Our final remarks are aimed at trying to give an overall perspective. Several
interesting features emerged in six dimensional gauge theories. For instance, 
properties of four dimensional gauge theories appear to have parallels in 
higher dimensions. One, which is not surprising, is that the Landau gauge 
anomalous dimension of the dimension two local gluon mass operator is the sum 
of the gluon and ghost anomalous dimensions. This follows purely as a 
consequence of the universal structure of BRST invariance and the dimension 
independent proof of \cite{104}. What was less apparent was the result for the 
electron wave function anomalous dimension. The gauge parameter dependence 
arose only in the one loop term in the six and eight dimensional cases to the 
various orders we computed. This may give some insight into the reasoning 
behind the four dimensional argument of \cite{108,109}. From another point of 
view it might be better to examine the connection of the field theories in 
different dimensions at a more fundamental level. A clue to this may be in the 
way we had to carry out our higher dimension renormalization. For instance, the
underlying master integrals were deduced using Tarasov's method, \cite{76,77}, 
which connects masters in $d$-dimensions to those in $(d+2)$-dimensions. While 
this is at a Feynman integral level there is a hint that there is a Lagrangian 
field theory connection which may be quantifiable using path integral methods. 
An indication of this here may be seen in the operators in various Lagrangians.
For instance, in (\ref{prop6m}) the mass parameters $m_2$ and $m_4$ are 
associated with $G_{\mu\nu}^a G^{a\,\mu\nu}$ and the Landau gauge operator 
${\cal O}$. In the corresponding four dimensional propagator, \cite{89,90,101},
the respective operators are ${\cal O}_2$ and the Gribov operator, 
${\cal O}_\gamma$, which is
\begin{equation}
{\cal O}_\gamma ~=~ \frac{1}{2} A^a_\mu \left( \frac{1}{\partial^\nu D_\nu}
\right)^{ab} A^b_\mu
\end{equation}
which is also nonlocal. The anomalous dimension of the latter is formally the 
same as ${\cal O}$ in that it is the sum of the gluon and ghost anomalous 
dimensions. However, comparing the structure of the respective operators
between four and six dimensions they are essentially equivalent when one
recognizes that the nonlocality accounts for the differing dimensionalities. 
This may be an indication that nonlocal problems in lower dimensions could be 
studied in a local higher dimensional context and give insight into effective 
field theories. 

\vspace{1cm}
\noindent
{\bf Acknowledgements.} The author thanks H. Kissler, D. Kreimer and R. Simms
for discussions. This work was carried out with the support of STFC 
Consolidated Grant ST/L000431/1.

\appendix

\section{Eight dimensional master integrals.}

In this appendix we record the values of the various one and two loop $3$-point
master integrals at the fully symmetric point needed to carry out the 
renormalization of (\ref{lag8}) and (\ref{lagqed8}) in eight dimensions. They 
are constructed from lower dimensional masters using Tarasov's method, 
\cite{76,77}. For ease of comparison and definition we use the same labelling 
of the integrals as that given in the four dimensional summary of \cite{110}. 
First, the one loop triangle integral is 
\begin{eqnarray}
{\cal M}^{(1)}_{31} &=& 
\left[ -~ \frac{1}{8\epsilon}
- \frac{61}{144} - \frac{2}{81} \pi^2 + \frac{1}{27} \psi^\prime(\third)
\right. \nonumber \\
&& \left. ~
+~ \left[ - \frac{895}{864} - \frac{23}{864} \pi^2 - \frac{2}{3} s_3(\pisix) 
+ \frac{1}{18} \psi^\prime(\third)
+ \frac{35}{5832} \sqrt{3} \pi^3 + \frac{1}{216} \sqrt{3} \ln^2(3) \pi 
\right] \epsilon \right. \nonumber \\
&& \left. ~+~ O(\epsilon^2) \frac{}{} \right] (-\tilde{\mu}^2) ~.
\end{eqnarray} 
At two loops we have 
\begin{eqnarray}
{\cal M}^{(2)}_{42} &=& \left[ \frac{1}{50400 \epsilon^2}
+ \frac{323}{4233600 \epsilon} 
+ \left[ 2234400 \psi^\prime(\third) - 1754200 \pi^2 + 6391701 \right]
\frac{1}{80015040000} \right. \nonumber \\
&& \left. ~
+ \left[ 81496800 \sqrt{3} \ln^2(3) \pi - 133358400 \sqrt{3} \ln(3) \pi
\right. \right. \nonumber \\
&& \left. \left. ~~~~~
+ 79301600 \sqrt{3} \pi^3 + 3775312800 \psi^\prime(\third)
+ 4800902400 s_2(\pisix) - 9601804800 s_2(\pitwo)
\right. \right. \nonumber \\
&& \left. \left. ~~~~~
- 18136742400 s_3(\pisix) + 6401203200 s_3(\pitwo) - 2773272600 \pi^2
- 2667168000 \zeta_3 
\right. \right. \nonumber \\
&& \left. \left. ~~~~~
- 11102348079 \right] \frac{\epsilon}{20163790080000} ~+~ O(\epsilon^2) 
\right] \tilde{\mu}^8 \nonumber \\
{\cal M}^{(2)}_{43} &=& \left[ \frac{1}{2880 \epsilon^2}
+ \frac{401}{172800 \epsilon}
+ \left[ - 2400 \psi^\prime(\third) - 3800 \pi^2 + 937449 \right]
\frac{1}{93312000}
\right. \nonumber \\
&& \left. ~
+ \left[ 108000 \sqrt{3} \ln^2(3) \pi - 1944000 \sqrt{3} \ln(3) \pi
- 244000 \sqrt{3} \pi^3 + 4600800 \psi^\prime(\third) 
\right. \right. \nonumber \\
&& \left. \left. ~~~~~
+ 69984000 s_2(\pisix)
- 139968000 s_2(\pitwo) - 108864000 s_3(\pisix) + 93312000 s_3(\pitwo)
\right. \right. \nonumber \\
&& \left. \left. ~~~~~
- 9563400 \pi^2 - 38880000 \zeta_3 + 611480367 \right]
\frac{\epsilon}{16796160000} ~+~ O(\epsilon^2) \right] \tilde{\mu}^6 
\nonumber \\
{\cal M}^{(2)}_{52} &=&
\left[ \frac{1}{960 \epsilon^2} + \frac{2371}{345600 \epsilon}
+ \left[ - 38400 \psi^\prime(\third) - 6800 \pi^2 + 5299929 \right]
\frac{1}{186624000} ~+~ O(\epsilon) \right] \tilde{\mu}^6 \nonumber \\
{\cal M}^{(2)}_{61} &=& \left[ \frac{1}{240 \epsilon^2}
+ \frac{329}{9600 \epsilon}
\right. \nonumber \\
&& \left. ~
+ \left[ 16200 \sqrt{3} \ln^2(3) \pi - 194400 \sqrt{3} \ln(3) \pi
- 17400 \sqrt{3} \pi^3 + 628800 \psi^\prime(\third) 
\right. \right. \nonumber \\
&& \left. \left. ~~~~~
+ 6998400 s_2(\pisix) - 13996800 s_2(\pitwo) - 11664000 s_3(\pisix) 
+ 9331200 s_3(\pitwo) 
\right. \right. \nonumber \\
&& \left. \left. ~~~~~
- 494800 \pi^2 - 1166400 \zeta_3 + 19175391 \right] \frac{1}{108864000} ~+~ 
O(\epsilon) \right] \tilde{\mu}^4 
\end{eqnarray}
where 
\begin{equation}
s_n(z) ~=~ \frac{1}{\sqrt{3}} \Im \left[ \mbox{Li}_n \left(
\frac{e^{iz}}{\sqrt{3}} \right) \right] 
\end{equation}
and $\mbox{Li}_n(z)$ is the polylogarithm function. We have used the notation
of \cite{110} but it is worth noting that they are related to cyclotomic
polynomials, \cite{111}. We have not included values for the two loop masters 
${\cal M}^{(1)}_{21}$, ${\cal M}^{(2)}_{31}$, ${\cal M}^{(2)}_{41}$ and 
${\cal M}^{(2)}_{51}$ in the notation of \cite{109} as they are products of one
loop masters or two loop $2$-point integrals.

Finally we record the value of the eight dimensional one loop $4$-point box 
integral at the fully symmetric point. This was required for the 
renormalization of (\ref{lag8}). In \cite{112} the four dimensional version was
derived but again we have used \cite{76,77} for our purposes. Using the same 
notation as \cite{112} the corresponding $d$~$=$~$8$~$-$~$2\epsilon$ 
dimensional value is
\begin{eqnarray}
D^{(1)} \left(-\tilde{\mu}^2,-\tilde{\mu}^2,-\tilde{\mu}^2,-\tilde{\mu}^2,
-\frac{4}{3}\tilde{\mu}^2,-\frac{4}{3}\tilde{\mu}^2\right) &=& 
\frac{1}{6\epsilon} + \frac{11}{18} 
- \frac{1}{24} \ln \left( \frac{4}{3} \right) 
+ \frac{25}{192} \Phi_1 \left( \frac{9}{16}, \frac{9}{16} \right) \nonumber \\
&& -~ \frac{29}{96} \Phi_1 \left( \frac{3}{4}, \frac{3}{4} \right) ~+~ 
O(\epsilon) 
\end{eqnarray}
where, \cite{112},
\begin{equation}
\Phi_1(x,y) ~=~ \frac{1}{\lambda} \left[ 2 \mbox{Li}_2(-\rho x)
+ 2 \mbox{Li}_2(-\rho y)
+ \ln \left( \frac{y}{x} \right)
\ln \left( \frac{(1+\rho y)}{(1+\rho x)} \right)
+ \ln(\rho x) \ln(\rho y) + \frac{\pi^2}{3} \right]
\end{equation}
and
\begin{equation}
\lambda(x,y) ~=~ \sqrt{\Delta_G} ~~~,~~~
\rho(x,y) ~=~ \frac{2}{[1-x-y+\lambda(x,y)]}
\end{equation}
with
\begin{equation}
\Delta_G(x,y) ~=~ x^2 ~-~ 2 x y ~+~ y^2 ~-~ 2 x ~-~ 2 y ~+~ 1 ~.
\end{equation}
We note that the finite piece can also be expressed in terms of the Clausen 
function $\mbox{Cl}_2(\theta)$ via, \cite{113},  
\begin{eqnarray}
\Phi_1 \left( \frac{3}{4},\frac{3}{4} \right) &=& \sqrt{2} \left[
2 \mbox{Cl}_2 \left( 2 \cos^{-1} \left( \frac{1}{\sqrt{3}} \right) \right)
+ \mbox{Cl}_2 \left( 2 \cos^{-1} \left( \frac{1}{3} \right) \right) \right]
\nonumber \\
\Phi_1 \left( \frac{9}{16},\frac{9}{16} \right) &=& \frac{4}{\sqrt{5}} \left[
2 \mbox{Cl}_2 \left( 2 \cos^{-1} \left( \frac{2}{3} \right) \right)
+ \mbox{Cl}_2 \left( 2 \cos^{-1} \left( \frac{1}{9} \right) \right) \right] ~.
\end{eqnarray}
The finite part has been provided for the reader interested in the Tarasov 
approach.

\end{document}